\documentclass[12pt]{article}

\usepackage{amsmath, amssymb, slashed}
\usepackage[dvips]{graphicx}
\usepackage{epsf}
\usepackage{color}

\begin{document}

\begin{titlepage}
\begin{center}

\hfill IPMU-12-0066\\
\hfill KEK-TH-1538\\
\hfill \today

\vspace{0.5cm}
{\large\bf Search for the Top Partner at the LHC \\
using Multi-b-Jet Channels}
\vspace{2.0cm}

{\bf Keisuke Harigaya}$^{(a, b)}$, 
{\bf Shigeki Matsumoto}$^{(a)}$,
\\
{\bf Mihoko M. Nojiri}$^{(c,d,a)}$
and 
{\bf Kohsaku Tobioka}$^{(a,b)}$

\vspace{0.5cm}
{\it
$^{(a)}${Kavli Institute of the Physics and Mathematics of the Universe (Kavli IPMU), \\TODIAS, University of Tokyo, Kashiwa, 277-8583, Japan} \\
$^{(b)}${Department of Physics, University of Tokyo, Bunkyo-ku 113-0033, Japan} \\
$^{(c)}${\it KEK Theory Center, Tsukuba, Ibaraki 305-0801, Japan}\\
$^{(d)}${\it The Graduate University for Advanced Studies (Sokendai), Department of Particle and Nuclear Physics, Tsukuba, Ibaraki 305-0801, Japan }
}
\vspace{2.0cm}

\abstract{Vector-like quarks are introduced in various new physics
 models beyond the standard model (SM) at the TeV scale. We especially
 consider the case that the quark is singlet (triplet) under the
 SU(2)$_L$ (SU(3)$_c$) gauge group and couples only to the third
 generation quarks of the SM. The vector-like quark of this kind is
 often called a top partner. The top partoner  $t_p$ decays into $bW,
 tZ$ and $th$. In the ATLAS and CMS collaborations, the top partner has
 been searched in the final states of $bW$ and $tZ$, while the search
 based on the decay mode $t_p\to th$ has not been started yet.  However,
 the decay into $th$ is important since it is significantly enhanced if
 some strong dynamics exists in the TeV scale. In the presence of a
 light higgs boson, the decay mode $t_p\to th$ followed by $h\to
 b\bar{b}$ produces three bottom quarks. We study the sensitivity for
 the top partner using multi-b-jet events at the 8 TeV run of the LHC
 experiment. The multi-b-jet eventss turn out to play a complementary
 role to the existing $t_p\rightarrow bW$ and $tZ$ searches by the ATLAS and CMS collaborations.}
\end{center}
\end{titlepage}
\setcounter{footnote}{0}

\section{Introduction}

Many new physics models at the TeV scale have been proposed so far to solve the hierarchy problem of the standard model (SM). Vector-like quarks, which are singlet (triplet) under the SU(2)$_L$ (SU(3)$_c$) gauge group, are often introduced in those models. One of famous examples is the little higgs model~\cite{LH} which has been proposed to solve the little hierarchy problem~\cite{little hierarchy}. In this model, the higgs boson is regarded as a pseudo-NB boson associated with a spontaneous symmetry breaking at 10 TeV. Explicit breaking terms are arranged to cancel quadratically divergent corrections to the higgs mass term at the 1-loop level, which stabilizes the higgs mass at ${\cal O}$(100) GeV. This arrangement is called the collective symmetry breaking, and it requires new vector-like quarks. The mass of the vector-like quark is predicted to be less than 600 (900) GeV to realize the fine-tuning less than 10\% (5\%) level~\cite{Harigaya:2011yg}, which is within the reach of the Large Hadron Collider (LHC) experiment.

Another example is the extension of the minimum supersymmetric standard model (MSSM) by introducing vector-like matters~\cite{vector matter}. This extension recently attracts attention because of the latest results of the LHC experiment, where the higgs mass is suggested to be 124--126GeV by both the ATLAS~\cite{Collaboration:2012si} and CMS~\cite{Chatrchyan:2012tx} collaborations. In the framework of the MSSM, it is difficult to achieve such high mass higgs boson for $M_{\rm SUSY} \sim$ 1 TeV~\cite{MSSM higgs mass}, leading to a little hierarchy problem. If there are new vector-like matters in the model, the higgs mass can be as heavy as 125 GeV while keeping $M_{\rm SUSY} \sim$ 1 TeV. Introduction of vector-like matters therefor allows us to set $M_{\rm SUSY}$ as light as current experimental limits which are, for example, consistent with results of the anomalous magnetic dipole-moment of the muon~\cite{MSSM higgs mass and g-2}.

In this article, we focus on the up-type vector-like quark whose
hypercharge is 2/3. We also postulate that the vector-like quark is
interacting only with the third generation quarks of the SM in order to
satisfy severe constraints from flavor changing
processes~\cite{FCNC}. The vector-like quark of this kind is often
called a top partner, and the character $t_p$ is used to denote it in
following discussions. In general, the top partner can decay into $bW, \
tZ, \ th, \ tg $ and $t\gamma$. The decay modes $t_p\to bW, tZ$ and $th$
are dominant because the other modes $t_p\to tg$ and $t_p \to t\gamma$
only appears in the 1-loop diagrams due to the gauge invariance. The top
partner is mixed with the SM top quark after the electroweak symmetry
breaking, which leads to the decays modes $t_p\to bW$ and $t_p\to tZ$. On the other hand, the decay into $th$ comes from the Yukawa interaction of the top partner and a dimension-five oparator. If some strong dynamics is the origin of new physics, such as in the little higgs model, the dimension-five operator can be sizable and the decay $t_p\to th$ is enhanced. 

The top partner decaying into $bW$ and $tZ$ have already been searched at the LHC experiment. For example, the CMS collaboration has searched for the process, $pp \to t_p\bar{t}_p X$ followed by the decay $t_p \to bW$, in a b-jet of high $p_T$ and lepton(s) channels with the integrated luminosity of 4.7 fb$^{-1}$. They have put a limit $m_{tp} >$ 560 GeV with Br($t_p \to bW$) = 1~\cite{CMS-bWbW2lep}, where $m_{tp}$ is the top partner mass. The CMS collaboration has also searched for another process, $pp \to t_p\bar{t}_p X$ followed by the decay $t_p \to tZ$, in multi-jet and three-lepton channel with 1.1 fb$^{-1}$ data. They have put a limit $m_{tp} >$ 475 GeV with Br($t_p \to tZ$) = 1~\cite{Chatrchyan:2011ay}. 

Experimental results  based on $t_p\to th$ have not been reported yet, although this is very important to  know the origin of the top partner.  The recent LHC data indicates the higgs mass of about 125 GeV \cite{Collaboration:2012si, Chatrchyan:2012tx}, and the higgs boson mostly decays into $b\bar{b}$ in this mass region. Since the decay mode $t_p\to th$  produces three bottom quarks in the presence of a such light higgs boson, multi-b-jet channels are expected to be sensitive for the detection of the top partner. In order to quantitatively investigate this expectation, we perform simulations including detector effects, and estimate the sensitivity of the LHC experiment to the top partner using multi-b-jet channels at the center of mass energy of 8 TeV. For comparison, we also estimate the sensitivity of the LHC experiment in the one b-jet and one lepton channel which has been adopted in the ATLAS and CMS collaborations.

There are previous studies that considerer the decay $t_p \to th$
\cite{T-to-th1, T-to-th2, T-to-th3}. Especially, one of them investigates
the top partner assuming the light higgs boson based on more than three
b-jets at the 14 TeV run of the LHC experiment \cite{T-to-th1}, and
another one studies the $t_p\to th$ followed by the decay of $h\to \gamma\gamma$ and $ZZ$\cite{T-to-th2} at 7 TeV and 14 TeV. 

We start with, in the next section, the effective action describing the top
partner in order to make the discussion as general as possible.  In
Section 3, we describe our simulation setup and perform the analyses in
the multi-b-jet channels, and show the sensitivity of the LHC experiment at the 8 TeV run.  According to results obtained in these analyses, we discuss some implications to new physics models beyond the SM in Section \ref{sec: application}. Section \ref{sec: summary} is devoted to summary of our discussions.

\section{Effective action for the top partner}
In order for our discussion to be as general as possible, we give an effective action of the top partners up to dimension-five operators. It is shown that a dimension-five operator actually takes an important role to enhance the branching fraction of the decay mode $t_p \to th$  which leads multiple b-jets at the LHC experiment. For more detail of this effective action, see Ref.\cite{Harigaya:2011yg}.
Since the top partner has quantum numbers of $({\bf 3}, {\bf 1}, 2/3)$ under the SM gauge groups, SU(3)$_c$ $\times$ SU(2)$_L$ $\times$ U(1)$_Y$, its interactions with higgs boson and third generation quarks of the SM are given by
\begin{eqnarray}
{\cal L}_{\rm eff}
=
- m_U \bar{U}_L U_R
- y_3 \bar{Q}_{3L} H^c u_{3R}
- y_U \bar{Q}_{3L} H^c U_R
- (\lambda/\Lambda) \, \bar{U}_L u_{3R} |H|^2
+ h.c.,
\label{eq: effective lagrangian}
\end{eqnarray}
where $H$, $Q_{3L}$, and $u_{3R}$ are higgs doublet, third generation left- and right-handed quarks, respectively, while $U_L$ and $U_R$ are left- and right-handed components of the top partner. The superscript `$c$' denotes charge conjugation, and $\Lambda$ in front of the dimension-five operator, $\bar{U}_L u_{3R} |H|^2$, is the cutoff scale, where the above effective action can be applied to describe physics below this scale. The other dimension-five operators such as $(\bar{U}_L U_R |H|^2)/\Lambda$ are  irrelevant  to our discussion. The top partner has the QCD interaction in addition to those in Eq.(\ref{eq: effective lagrangian}). All the parameters in the effective action can be real by appropriate redefinitions of the fields. Note that there are only three free parameters because the top quark mass $m_t$ has already been measured.  The use of this effective action is particularly useful for the little higgs model, because the top partner is nothing but a new particle which is introduced to cancel the quadratically divergent correction to the higgs mass term from the top loop diagram, and therefore the top partner mass should be lighter than the other new particles. The effective action is valid only when the other new particles are heavy enough compared to the top partner.

After the electroweak symmetry is broken down, left- and right-handed components of the top partner ($U_L$ and $U_R$) are mixed with those of the SM top quark,
\begin{eqnarray}
\begin{pmatrix} t_L \\ {t_p}_L \\ \end{pmatrix}
=
\begin{pmatrix}
\cos \theta_{tL} & -\sin \theta_{tL} \\
\sin \theta_{tL} &  \cos \theta_{tL} \\
\end{pmatrix}
\begin{pmatrix} u_{3L} \\ U_L \\ \end{pmatrix},
~~
\begin{pmatrix} t_R \\ {t_p}_R \\ \end{pmatrix}
=
\begin{pmatrix}
\cos \theta_{tR} & -\sin \theta_{tR} \\
\sin \theta_{tR} &  \cos \theta_{tR} \\
\end{pmatrix}
\begin{pmatrix} u_{3R} \\ U_R \\ \end{pmatrix},
\label{eq: mixing}
\end{eqnarray}
where these mixing matrices diagonalize the mass matrix of the top partner and the top quark, giving their mass eigenvalues $m_{tp}$ and $m_t$. In the following, we take $m_{tp}$, $m_t$, and $\sin \theta_{tL}$, and $\sin \theta_{tR}$ as model parameters instead of $m_U$, $y_3$, $y_U$, and $\lambda/\Lambda$ which are originally defining the effective action. With $s_{tL}$ ($c_{tL}$) and $s_{tR}$ ($c_{tR}$) being $\sin \theta_{tL}$ ($\cos \theta_{tL}$) and $\sin \theta_{tR}$ ($\cos \theta_{tR}$), respectively, the original parameters are given
\begin{eqnarray}
m_U &=& s_{tR}s_{tL} m_t - c_{tR}c_{tL}m_{tp}, \\
y_{3} &=& (\sqrt{2}/v)(c_{tR}c_{tL}m_t + s_{tR}s_{tL}m_{tp}), \\
y_U &=& (\sqrt{2}/v)(-s_{tR}c_{tL}m_t + c_{tR}s_{tL}m_{tp}), \\
\lambda/\Lambda &=& (2/v^2)(-c_{tR}s_{tL}m_t + s_{tR}c_{tL}m_{tp}),\label{lambda}
\end{eqnarray}
where $v$ is the vacuum expectation value of the higgs field. All physical quantities related to the LHC signal of the top partner are obtained with the use of this parameterization as well as gauge interactions of top partner and top quark.

\begin{figure}[t]
\begin{center}
\includegraphics[scale=0.2]{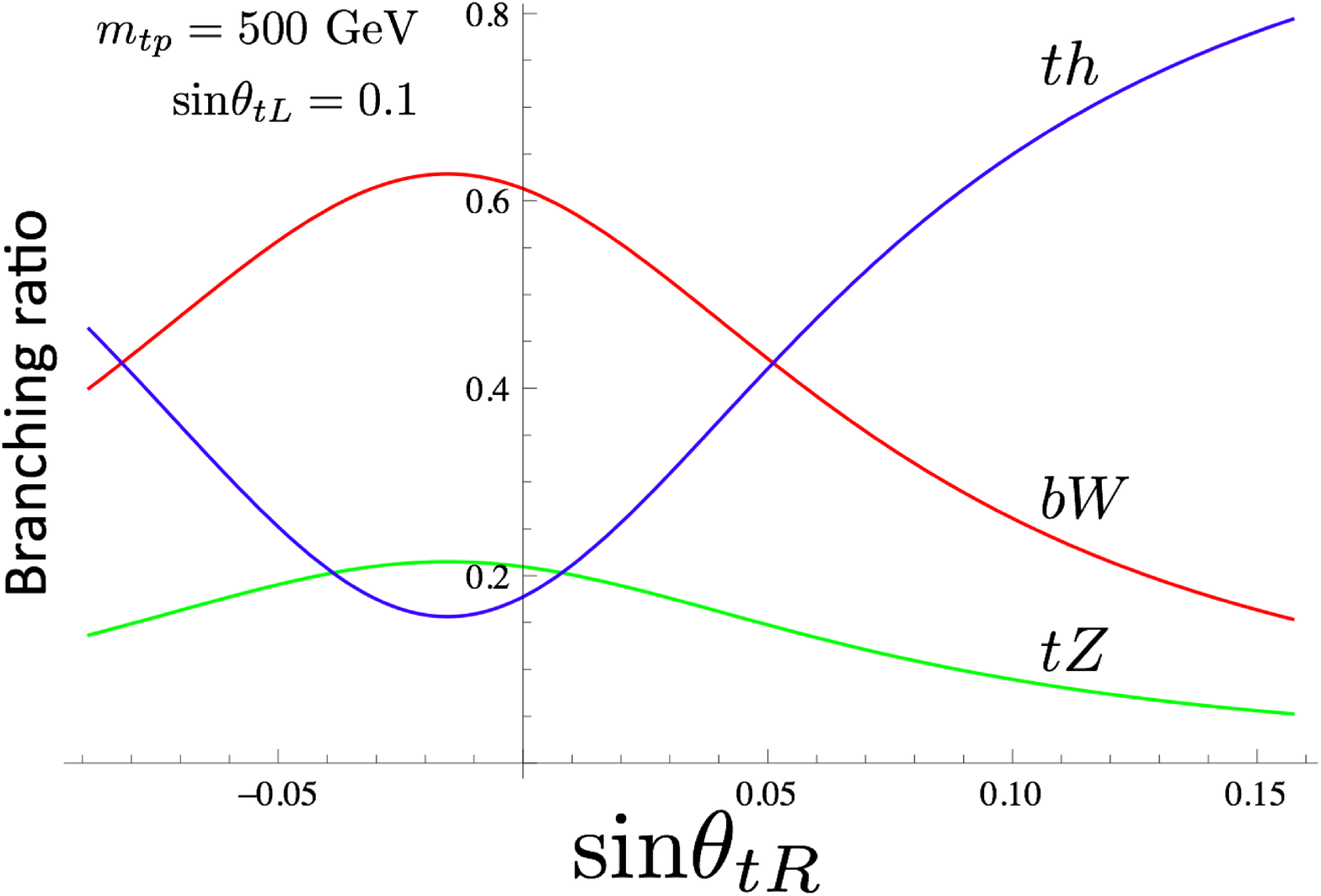}
\caption{\small The branching fraction of the decay of the top partner (${t_p} \to bW$, $tZ$, and $th$) as a function of $\sin \theta_{tR}$ with $\sin \theta_{tL}$= 0.1 and $m_{t_p}$= 500 GeV.}
\label{fig: branching ratio}
\end{center}
\end{figure}

In Fig.\ref{fig: branching ratio}, the branching fraction of the decay
of the top partner is depicted as a function of $\sin \theta_{tR}$ with
$\sin \theta_{tL}=0.1$ and $m_{t_p}$ = 500 GeV. As can be seen in
Section 4, the left-handed mixing angle $\sin \theta_{tL}$ is severely
constrained by the electroweak precision measurement and thus should be
small enough. Since branching fractions of decay modes $t_p \to bW$ and
$t_p \to tZ$ are proportional to $\sin^2 \theta_{tL}$, the decay mode
$t_p \to th$ has a large branching fraction when $\sin \theta_{tR}$ is
sizable. Since the first term of Eq.(\ref{lambda}) is negligible
compared to the second term, the right-handed mixing angle $\sin
\theta_{tR}$ depends strongly on the coefficient of the dimension-five
operator $\lambda/\Lambda$. In the little higgs model, $\Lambda/\lambda$ is of
the order of the top partner mass $m_{tp}$, and, as a result, $\sin \theta_{tR}$ can be as large as 0.1. In such a case, the branching fraction of the decay mode ${t_p} \to th$ is enhanced and the use of the multi-b-jet channel becomes efficient to search for the top partner. When the scale $\Lambda$ is large enough as in the most cases of weak-interacting new physics models, the angle $\sin \theta_{tR}$ is almost zero. 

\section{Multi-b-jet channel}
\label{sec: simulation}

In this section, we study the top partner signature in multi-b-jet channels and the SM backgrounds. After describing our simulation framework and selection-criteria to reduce the backgrounds, we present the sensitivity of the LHC experiment to the top partner signal in not only multi-b-jet channels but also the channel used in previous analyses of the ATLAS and CMS collaborations. The center of mass energy is set to be 8 TeV and higgs mass is 120 GeV throughout this article.

\subsection{Simulation framework}
\label{sec: setup}

Using the Feynrules~\cite{Feynrule} package, we first implement the interactions of $t_p$ into MadGraph5~\cite{Alwall:2011uj} based on the effective action. Parton level events are interfaced to PYTHIA6.420~\cite{Sjostrand:2006za} for parton-showering and hadronization, and Delphes1.9~\cite{Ovyn:2009tx} is used to simulate detector effects. We set appropriate resolution-parameters for the detector simulation based on the ATLAS detector performance~\cite{ATLAS-detector}. 

We adopt the method to reconstruct objects such as isolated, central leptons and
jets according to the strategy of new physics searches at the
ATLAS experiment~\cite{ATLAS:2011ad}.  Jet candidates are reconstructed
using the anti-$k_T$ algorithm~\cite{anti-kt} implemented in Delphes1.9
with the radius parameter $R = 0.4$. The jet candidates are required to have
the transverse
momentum $p_T>20$ GeV and the pseudo rapidity $|\eta|<2.8$.
Electron and muon candidates are
identified via the generator-data assuming 100\%
efficiency. The electron(muon) candidates are required to have $p_T>20(10)$ GeV
and $|\eta|<2.47(2.4)$.
After these pre-selections, overlaps between the electron candidates and
the jet candidates are removed. The jet candidates are discarded if
their distances $\Delta R=\sqrt{\Delta\eta^2+\Delta \phi^2}$ to any electron
candidates are less than 0.2 where $\Delta \eta(\Delta \phi)$ is the
difference of the pseudo rapidity (azimuthal angle) between the jet
candidate and the electron candidate. The remaining jet candidates are
called ``jets''. For each jets, the electron candidates with $0.2<\Delta
R<0.4$ from jets are removed. Furthermore, isolation criteria are
imposed.
The electron candidates are removed if the scalar sums of the transverse
momentum of tracks within a cone size of $\Delta R=0.2$ around the
electron candidates exceed 10\% of the electron candidates.
The muon candidates are removed if the scalar sums of the transverse
momentum of tracks within a cone size of $\Delta R=0.2$ around the
muon candidates exceed 1.8 GeV. The remaining electron(muon)
candidates are called ``electrons(muons)''. We also take a calibrated efficiency and a mis-tagging rate for the b-tagging obtained by the SVO50 method~\cite{ATLAS-b-tag}. These results are well fitted by
\begin{eqnarray}
({\rm b\mathchar`-tag \, efficiency})
&=&
0.6 \, \tanh(p_T/36 \, {\rm GeV}) \times (1.02 - 0.02 \, |\eta|),
\label{b-tagging}\\
({\rm mistag \, rate \, for \, light\mathchar`-jet)}
&=&
0.001 + 0.00005 \, (p_T/1 \, {\rm GeV}),
\end{eqnarray}
where $p_T$ and $\eta$ are transverse momentum and pseudo-rapidity of the jet, respectively. On the other hand, we use the value of 0.14 as the mistag rate for $c$-jet, which is obtained by the method of the secondary vertex~\cite{ATLAS:1999fq}. We find our detector simulation reproduces the ATLAS data involving a b-jet, leptons and missing energy within 10\% accuracy. The comparison with the ATLAS data is presented in appendix \ref{app: validation}. 

\begin{table}[t]
\begin{center}
\begin{tabular}{|c|cccccccc|}
\hline
Mass (GeV) & 400 & 450 & 500 & 550 & 600 & 650 & 700 & 750 \\
\hline
Cross section (pb) & 2.0 & 0.99 & 0.51 & 0.27 & 0.15 & 0.086 & 0.051 & 0.030 \\
\hline
\end{tabular}
\caption{\small Cross sections of $pp \to t_p\bar{t}_p + X$ with several choices of the top partner mass.}
\label{table: signal CS}
\end{center}
\end{table}

The top partner can be copiously produced unless it is too heavy. The
dominant process to search for the top partner is the pair production,
$pp \to t_p\bar{t}_p + X$. The production cross section obtained by
HATHOR~\cite{Aliev:2010zk} is shown in Table \ref{table: signal
CS}. From the effective action Eq.(1), the top partner has three decay
modes; $t_p \to$ $bW$, $tZ$, and $th$. In the previous study, the top
partner has been searched for in one b-jet  and one lepton channel
aiming for the $t_p \to bW$ decay. In this article, we study the decay
mode $t_p \to th$ which is sensitive to the cutoff scale of the little
higgs model, as mentioned  in Section 2. In order to search for the
decay, the signature with multiple b-jets is more important since one
$t_p$-decay can produce three b-jets.

There are several background processes against the signal, which are listed in Table \ref{table: generator}. In order to generate the $t\bar{t}+$jets events, we have considered processes associated with up to two additional partons ($t\bar{t}$, $t\bar{t}j$, and $t\bar{t}jj$). These events are matched after the parton-showering using the MLM matching scheme implemented in MadGraph. For the generation of the $W+$jets events, the $W$ boson is forced to decay leptonically and is associated with up to three additional partons, where the first and second partons are required to have $p_T >$ 70 GeV and 40 GeV, respectively.  For $t\bar{t} b\bar{b}$ events, the first b-parton is required to have $p_T >$ 30 GeV. 
The background conditions as well as their cross sections are also listed in the second and third columns of Table \ref{table: generator}. The total cross section of $t\bar{t}+$jets is normalized to be 204 pb, which is again obtained by HATHOR. For the others, we use the matched cross sections obtained from the MadGraph-PYTHIA package. 
There are other possible backgrounds such as $Z+$jets, $W b\bar{b}+$jets, $Z b\bar{b}+$jets and $b\bar{b}b\bar{b}$, but we have confirmed those are negligible in our analysis.

\begin{table}[t]
\begin{center}
\begin{tabular}{c|cc}
& Remarks & Cross section (pb) \\
\hline
$t\bar{t}+$jets & up to 2 jets & 204 \\
$W+$jets & up to 3 jets & 320 \\
& ($W \to l\nu$, \, $p_{T{\rm \mathchar`-1st, 2nd \, j}} >$ 70, 40GeV) & \\
$t\bar{t} b\bar{b}$ & ($p_{T{\rm \mathchar`-1st \, b}} >$ 30GeV) & 1.6 \\
\hline
\end{tabular}
\caption{\small Background processes against the signal of the top partner.}
\label{table: generator}
\end{center}
\end{table}

\subsection{Event selections}
\label{sec: selection}

\subsubsection{Single-b-jet channel: 1 lepton + $\geq$1b-jet}

The pair production of top partners followed by the decays into two
bottom quarks and two $W$ bosons has been searched for in the CMS
collaboration ~\cite{CMS-bWbW2lep}. The one-lepton channel, ${t_p}
\bar{t}_p \to b W^+ \bar{b} W^- \to b \bar{b} l^{\pm} \nu jj$, currently
gives the most stringent bound on the ${t_p}\bar{t}_p$ production. We
therefore impose the following selection-criteria in order to evaluate the LHC-sensitivity to the signal,
\begin{enumerate}
\item There is only one lepton which is required to have $p_T >$ 30 GeV.
\item Transverse missing energy $\slashed{E}_T$ should be larger than 20 GeV.
\item There are more than three jets, and at least one of them is b-tagged.
\item Leading four jets are required to have $p_T >$ 80, 50, 30, and 30 GeV.
\end{enumerate}
Energetic b-jets are expeted from any decay pettern of $t_p \bar{t}_p$.  
The distribution of invariant mass $M_{bl\nu}$ is very useful to
separate signal events from those of the SM backgrounds mostly from top
quarks. Here, the invariant mass $M_{bl\nu}$ is reconstructed from the
leading b-jet, a charged lepton detected, and missing transverse
momentum $P_{Tmiss}$. The transverse momentum of the neutrino is
identified with $P_{Tmiss}$, while the longitudinal one is determined so
that the $W$ boson mass is reproduced from the lepton and the
neutrino. In order to deal with the two-fold ambiguity, which appears in
the determination of the longitudinal momentum, we always take the
smaller $M_{bl\nu}$. If there is no solution of the longitudinal momentum
that satisfies $(p_{{\rm lepton}}+p_{{\rm miss}})^2=m_W^2$, we take the momentum so that it gives the minimal $M_{bl\nu}$. We then require following three more selection-criteria to generated signal and background events,
\begin{enumerate}
\setcounter{enumi}{4}
\item The leading b-jet is required to have $p_T >$ 260 GeV.
\item The invariant mass $M_{\rm bl\nu}$ should be larger than 400 GeV
\item The effective mass $M_{\rm eff}$ should be larger than 1000 GeV,
\end{enumerate}
where the effective mass in the above criteria is defined by $M_{\rm eff} =
p_{T\mathchar`-{\rm lepton}} + \slashed{E}_T + \sum_{i = 1}^{{\rm
min}(5,N_{{\rm jets}})} p_{T\mathchar`-i{\rm th \, j}}$. Here, $N_{\rm
jets}$ is the number of the jets in the events. These criteria enable us to reduce $t\bar{t}+$jets background significantly.

\begin{figure}[p]
\begin{center}
\includegraphics[scale=0.44]{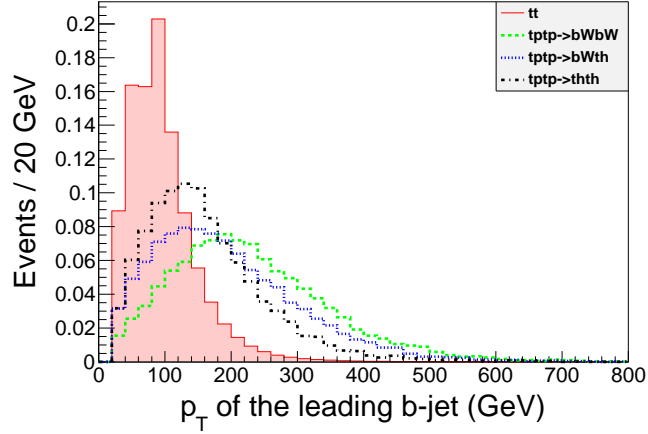} \\
\vspace{0.5cm}
\includegraphics[scale=0.44]{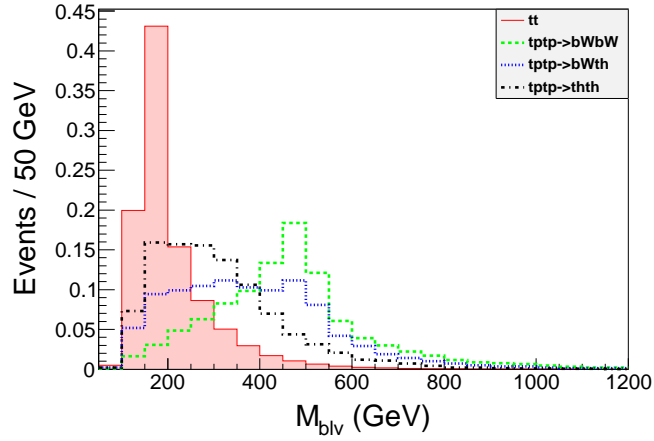} \\
\vspace{0.5cm}
\includegraphics[scale=0.44]{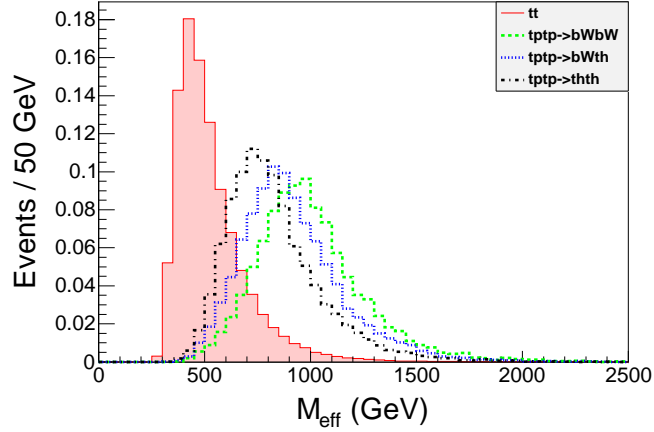}
\caption{\small Distributions of $p_{T\mathchar`-1{\rm st \, b}}$ (the transverse momentum of the leading b-jet), $M_{\rm bl\nu}$, and $M_{\rm eff}$ after applying kinematical cuts 1--4 in the analysis of the `1 lepton + $\geq$1b-jet' channel. The mass of the top partner is  $m_{tp} =$ 500 GeV. Distributions of $t_p \bar{t}_p \to$ $bWbW$, $bWth$ and $thth$ are shown separately assuming all $t_p \bar{t}_p$-decays in each mode. The distributions are normalized so that their integrated values become unity.}
\label{fig: 1b-dist}
\end{center}
\end{figure}

Distributions of $p_{T\mathchar`-1{\rm st \, b}}$ (the transverse
momentum of the leading b-jet), $M_{\rm bl\nu}$, and $M_{\rm eff}$ after
applying kinematical cuts 1--4 are plotted in Fig.\ref{fig: 1b-dist}
with $m_{tp} =$ 500 GeV. Distributions of $t_p \bar{t}_p \to$ $bWbW$,
$bWth$ and $thth$ are shown separately assuming all $t_p
\bar{t}_p$-decays in each mode. The distribution of the $t\bar{t}+$jets
events is also plotted in the same figure. The $p_T$-distribution of
b-jet for $t_p \bar{t}_p \to bWbW$ is highest among all the
distributions and has a clear peak at $M_{bl\nu} \sim m_{tp}$. The separation between signal and background is the worst for the $ t_p \bar{t}_p \to thth$ events. 

We present the cut flows of both signal and background events in  Table
\ref{table: 1b-cutflow}. Numbers of the signal and background events are
normalized to those corresponding to 15 fb$^{-1}$, which are shown in
the third low. In the last low of Table \ref{table: 1b-cutflow}, we give
acceptances after all the kinematical cuts imposed. Acceptances of the signal events for various masses of the top partner are also found in Table~\ref{table: 1b-acceptance}.  The $t_p \bar{t}_p \to bWth$ and $thth$ events are found to decrease after applying kinematical cuts 5 and 6.  The contribution of the $t_p \bar{t}_p \to bWth$ channel is larger when the top partner mass is heavier, because it is easier to satisfy kinematical cuts on $p_T$. 

For comparison, in appendix \ref{app: CMS-based analysis}, we also show the results based on the CMS-analysis method described in Ref.\cite{CMS1}. Our estimate gives a better result than that based on the CMS analysis. 

\begin{table}[t]
\begin{center}
{\small
\begin{tabular}{c|ccccc}
& $t\bar{t}$ & $W$+jets& $T\bar{T} \to bWbW$ & $T\bar{T} \to bWth$
& $T\bar{T} \to thth$ \\
\hline
Generated    & 6163292 & 809561  & 60000 & 60000 & 60000 \\
Without cuts & 3060000 & 4800000 & 7650  & 7650  & 7650 \\
Cuts 1--3    & 282861  & 16756   & 1336  & 1652  & 1693 \\ 
Cut 4        & 133099  & 6392    & 1160  & 1555  & 1654 \\
Cut 5        & 1872    & 415     & 405   & 374   & 215 \\
Cut 6        & 1025    & 320     & 351   & 287   & 125 \\
Cut 7        & 387     & 160     & 237   & 164   & 75 \\
\hline
Acceptance  & 0.00013  & 0.000033& 0.031 & 0.021 & 0.0099 \\
\hline
\end{tabular}
}
\caption{\small Cut flows of the signal and background events in the `1 lepton + $\geq$1b-jet' channel. The mass of the top partner is  $m_{tp} =$ 500 GeV. Results of $t_p \bar{t}_p \to$ $bWbW$, $bWth$ and $thth$ are shown separately assuming all $t_p \bar{t}_p$-decays in each mode.}
\label{table: 1b-cutflow}
\end{center}
\end{table}

\begin{table}[t]
\begin{center}
{\small
\begin{tabular}{c|cccccccc}
Mass (GeV) & 400 & 450 & 500 & 550 & 600 & 650 & 700 &750 \\
\hline
$T\bar{T} \to bWbW$ & 0.0096  & 0.018  & 0.031 & 0.047 & 0.062 & 0.074
			 & 0.084 & 0.092 \\
$T\bar{T} \to bWth$ & 0.0067 & 0.013  & 0.021 & 0.034 & 0.048 & 0.063
			 & 0.076 & 0.090 \\
$T\bar{T} \to thth$ & 0.0030 & 0.0050 & 0.0098 & 0.016 & 0.026 & 0.036
                         & 0.049 & 0.059 \\
\hline
\end{tabular}
}
\caption{\small
Acceptances of the signal events in the `1 lepton + $\geq $1b-jet' channel with several choices of the top partner mass. Results of $t_p \bar{t}_p \to$ $bWbW$, $bWth$ and $thth$ are shown separately assuming all $t_p \bar{t}_p$-decays in each mode.}
\label{table: 1b-acceptance}
\end{center}
\end{table}

\subsubsection{Two-b-jet channles: 1 lepton + $\geq$2b-jets}

We next consider the two-b-jet channels to search for the top partner. Both decay modes of the top partner, ${t_p} \to bW$ and ${t_p} \to th$, are expected to contribute to signal events in this channel. On the other hand, certain amounts of background events are also expected from $t\bar{t}+$jets processes. As in the case of the single-b-jet channel, we also focus on the two-b-jet channels associated with semi-leptonic decays of two $W$ bosons to reduce the SM backgrounds,
\begin{eqnarray}
PP \to {t_p}\bar{t}_p \to bWbW, \, \, bWth, \, \, thth
\to WW + {\rm b\mathchar`-jets}
\to l\nu jj + {\rm b\mathchar`-jets}.
\end{eqnarray}
Therefore, in the analysis of this channel, an efficient selection-criteria turns out to be essentially the same as that of the single-b-jet analysis, except for the number of b-jets in order to extract the signal events with multiple b-jets. With the use of exactly the same method for the reconstruction of the invariant mass $M_{bl\nu}$ adopted in the previous single-b-jet analysis, we impose the following selection-criteria
\begin{enumerate}
\item There is only one lepton which is required to have $p_T >$ 30 GeV.
\item Transverse missing energy $\slashed{E}_T$ should be larger than 20 GeV.
\item There are more than four jets, and at least two of them are b-tagged.
\item The b-jets are required to have $p_T >$ 200, 180 GeV.
\item The invariant mass $M_{bl\nu}$ should be larger than 250 GeV.
\item The effective mass $M_{\rm eff}$ should be larger than 1200 GeV.
\end{enumerate}

\begin{figure}[p]
\begin{center}
\includegraphics[scale=0.37]{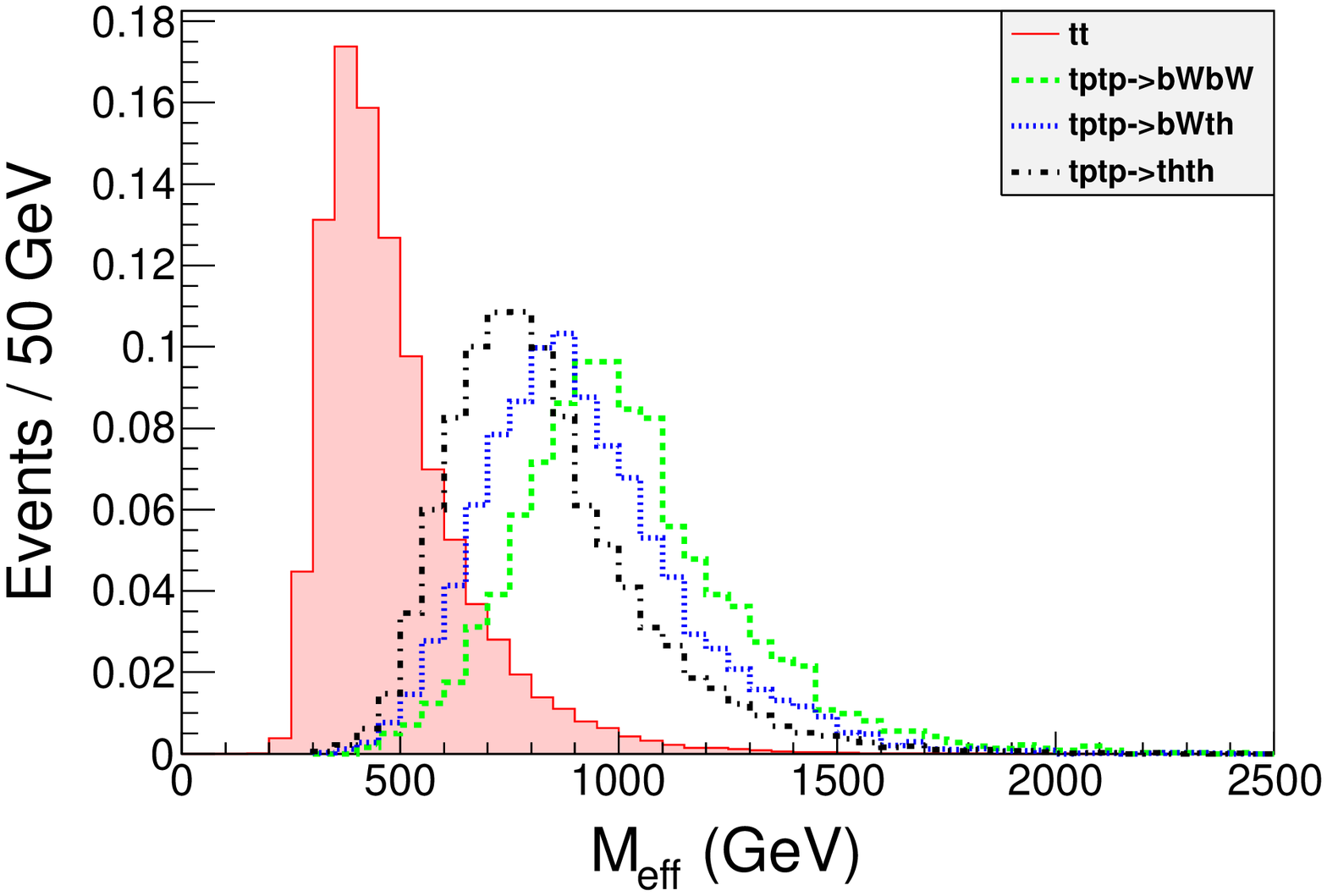}
\includegraphics[scale=0.37]{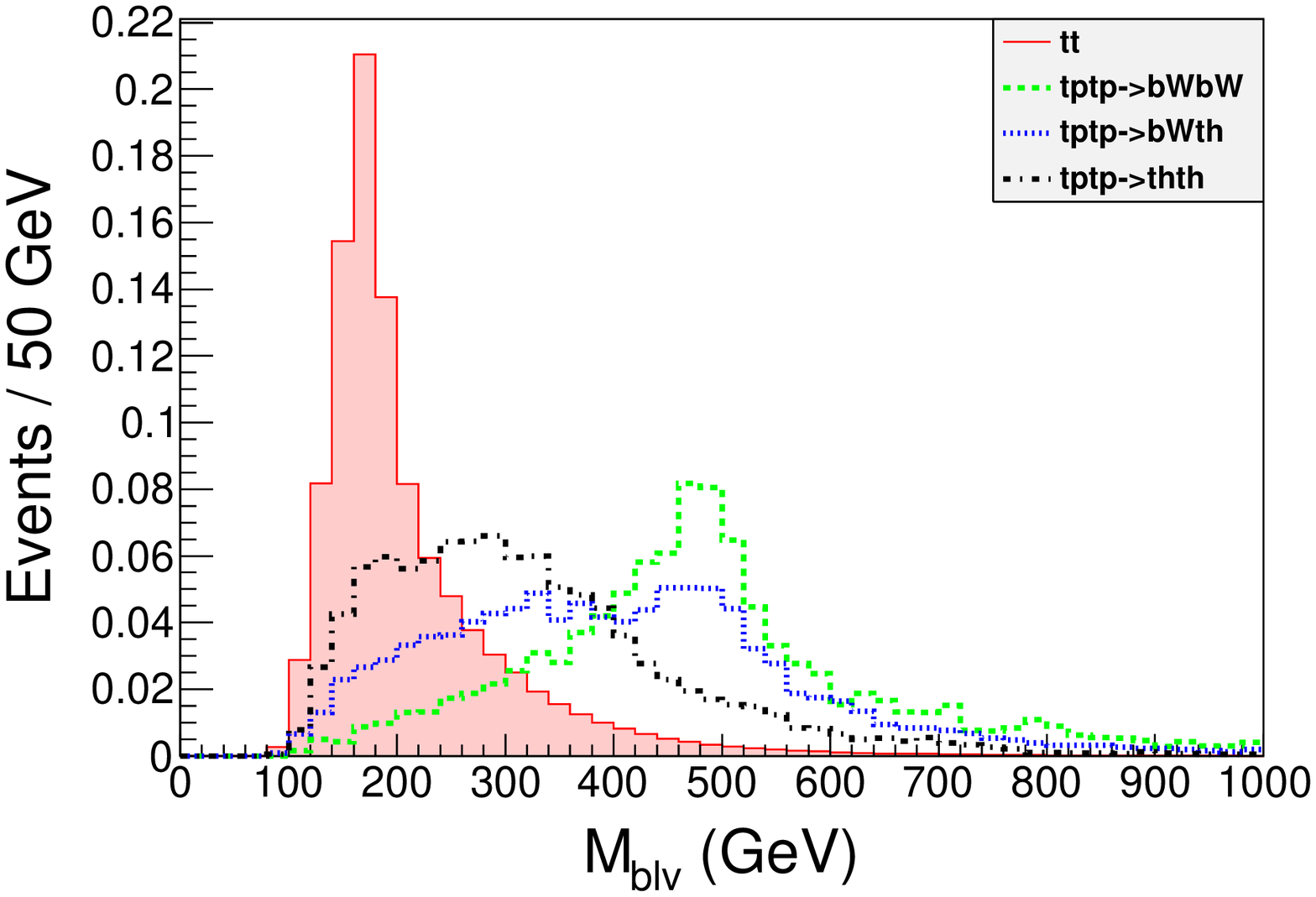} \\
\vspace{1.0cm}
\includegraphics[scale=0.37]{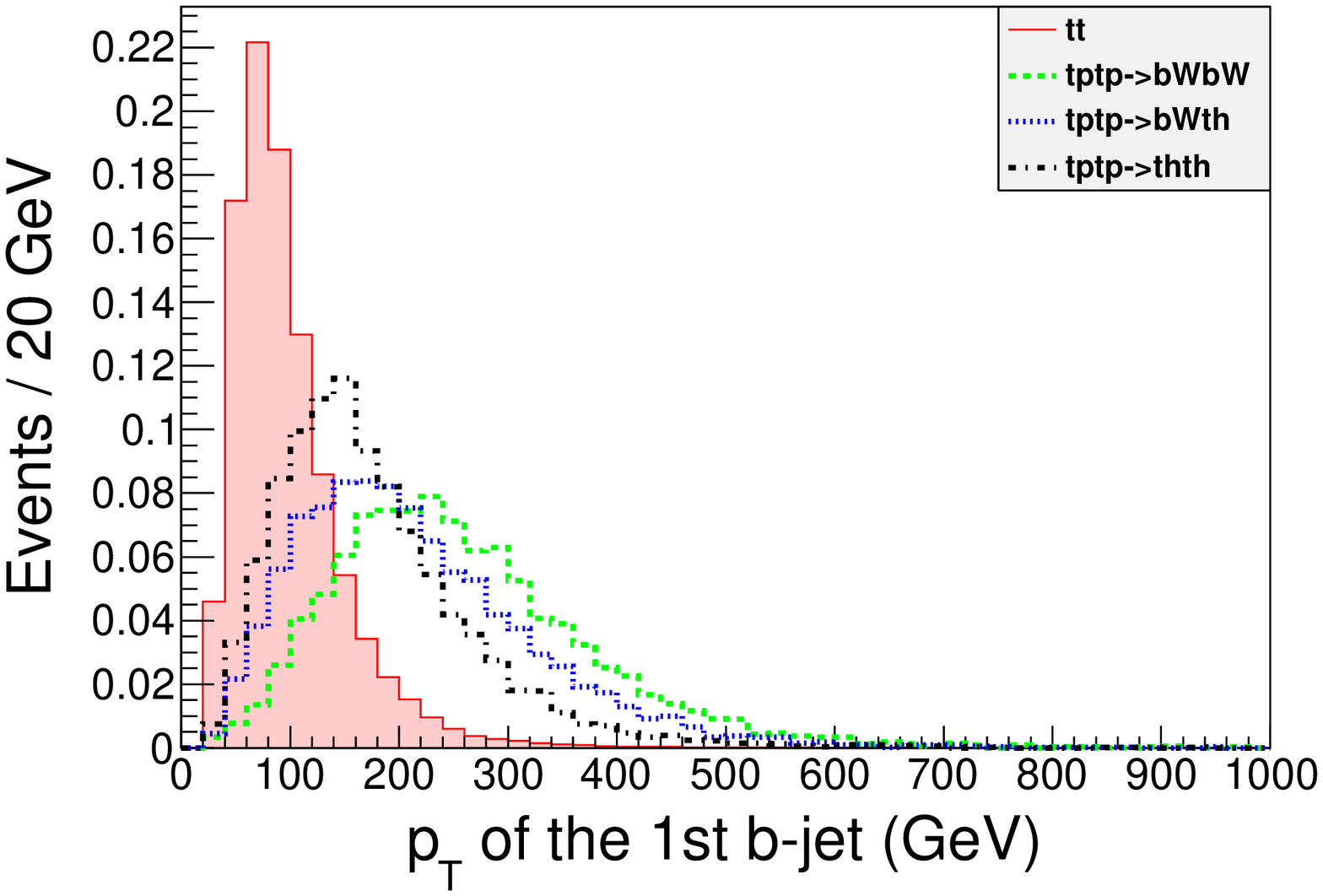}
\includegraphics[scale=0.37]{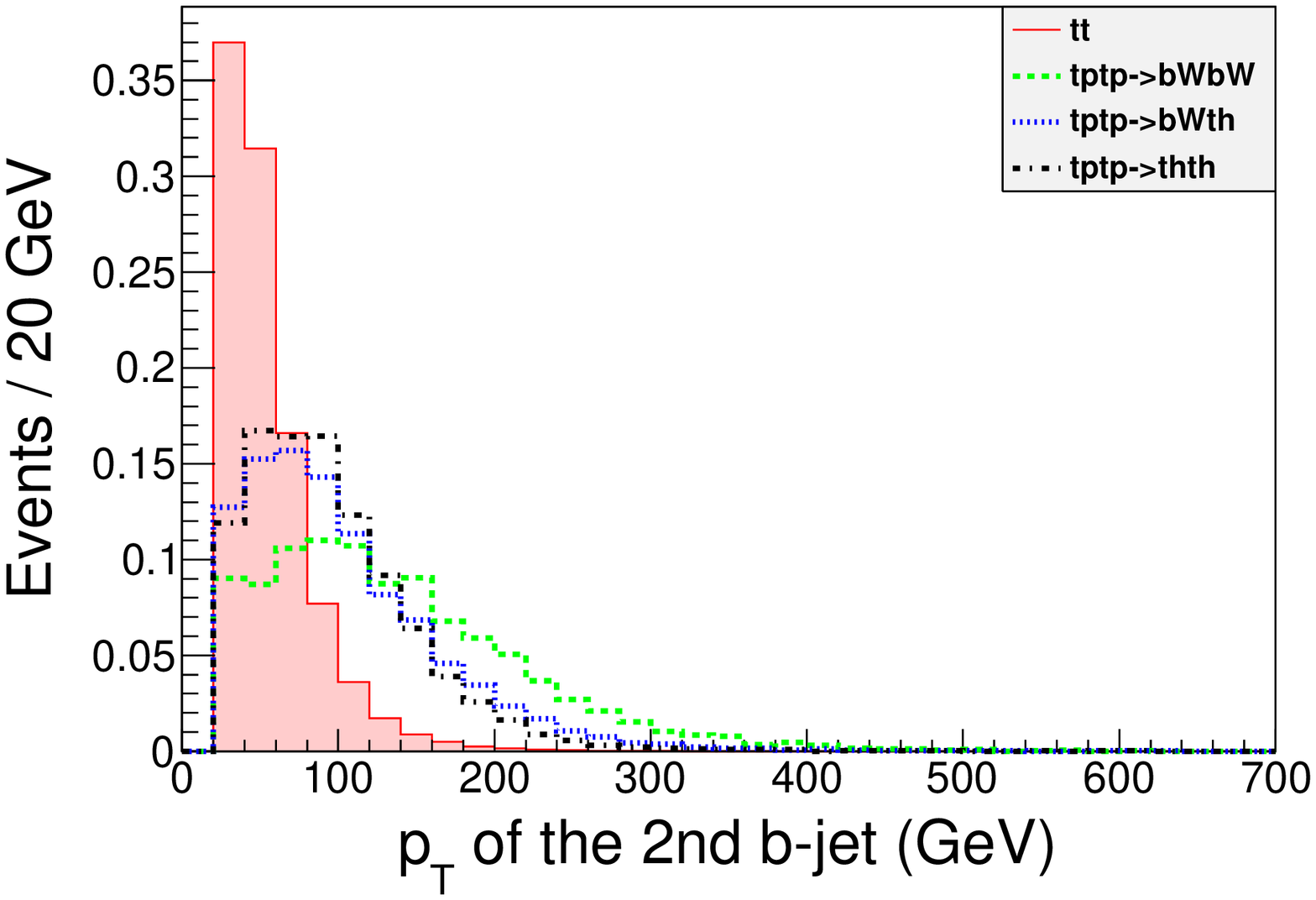}
\caption{\small Distributions of $M_{\rm eff}$, $M_{bl\nu}$, and transverse momenta of the first and second b-jets for signal and background ($t\bar{t}+$jets) events after applying the cuts 1--3 in the analysis of the `1 lepton + $\geq$2b-jets' channel. The mass of the top partner is  $m_{tp} =$ 500 GeV. Distributions of $t_p \bar{t}_p \to$ $bWbW$, $bWth$ and $thth$ are shown separately assuming all $t_p \bar{t}_p$-decays in each mode. The distributions are normalized so that their integrated values become unity.}
\label{fig: 2b-dist}
\end{center}
\end{figure}

Distributions of $M_{\rm eff}$, $M_{bl\nu}$, and transverse momenta of the first and second b-jets after applying kinematical cuts 1--3 are plotted in Fig.\ref{fig: 2b-dist}. As in the case of the previous single-b-jet analysis, the mass of the top partner for signal events is again  $m_{t_p} =$ 500 GeV, and we concider decay patterns of $t_p\bar{t}_p\to$ $bWbW$, $bWth$, and $thth$.  The distribution of $t\bar{t}+$jets events is plotted in the same figure. The cut flows of the signal events and  the main background events($t\bar{t}+$jets and $t\bar{t}b\bar{b}$) are shown in Table \ref{table: 2b-cutflow}. Numbers of events for signals and backgrounds in the third low correspond to those with the integrated luminosity of 15 fb$^{-1}$, while their acceptances after applying all kinematical cuts 1--6 are shown in the last low. The acceptances  in each decay pattern of $t_p\bar{t}_p$ are found in Table \ref{table: 2b-acceptance}.

\begin{table}[t]
\begin{center}
{\small
\begin{tabular}{c|ccccc}
& $t\bar{t}$ & $t\bar{t}b\bar{b}$ & $T\bar{T} \to bWbW$ & $T\bar{T} \to bWth$
& $T\bar{T} \to thth$ \\
\hline
Generated    & 6163292  & 90000   & 60000 & 60000 & 60000 \\
Without cuts & 3060000  & 24000   & 7650   & 7650   & 7650 \\
Cuts 1--3    & 56143    & 1677    & 413    & 936   & 1227 \\
Cut 4        & 317      & 17      & 103    & 100    & 79 \\
Cut 5        & 209      & 14      & 100    & 92     & 67 \\
Cut 6        & 33       & 3       & 37     & 30     & 20 \\
\hline
Acceptance   & 0.000011 & 0.00011 & 0.0049 & 0.0040 & 0.0026 \\
\hline
\end{tabular}
}
\caption{\small Cut flows of signal and background events in the analysis of the `1 lepton + $\geq$2b-jets' channel. The mass of the top partner is $m_{t_p} =$ 500 GeV. Results of $t_p \bar{t}_p \to$ $bWbW$, $bWth$ and $thth$ are shown separately assuming all $t_p \bar{t}_p$-decays in each mode.}
\label{table: 2b-cutflow}
\end{center}
\end{table}

\begin{table}[t]
\begin{center}
{\small
\begin{tabular}{c|cccccccc}
Mass (GeV) & 400 & 450 & 500 & 550 & 600 & 650 & 700 & 750 \\
\hline
$T\bar{T} \to bWbW$ & 0.0015 & 0.0026 & 0.0049 & 0.0075 & 0.012  & 0.016
			 & 0.020 & 0.024 \\
$T\bar{T} \to bWth$ & 0.0010 & 0.0023 & 0.0040 & 0.0065 & 0.011  & 0.016
			 & 0.022 & 0.028 \\
$T\bar{T} \to thth$ & 0.00077 & 0.0015 & 0.0026 & 0.0040 & 0.0064 & 0.011
			 & 0.015 & 0.022 \\
\hline\end{tabular}
}
\caption{\small Acceptances of signal events in the `1 lepton + $\geq $2b-jets' channel with several choices of the top partner mass. Results of $t_p \bar{t}_p \to$ $bWbW$, $bWth$ and $thth$ are shown separately assuming all $t_p \bar{t}_p$-decays in each mode.}
\label{table: 2b-acceptance}
\end{center}
\end{table}

\subsubsection{Three-b-jet channels: 1 lepton + $\geq$3b-jets}

We finally consider the three-b-jet channels to search for the top partner. This channel relies on the decay mode $t_p \to t h$. Since the higgs mass is expected to be about 120--130 GeV as strongly suggested by recent data of the ATLAS~\cite{Collaboration:2012si} and the CMS~\cite{Chatrchyan:2012tx} collaborations, the higgs boson decays mainly into two bottom quarks, and, as a result, the decay of the top partner produces three bottom quarks. Several three-b-jet channels following the pair production of the top partner are, in fact, available. Among those, in order to reduce SM backgrounds efficiently, we again focus on the channel associated with semi-leptonic decays of two $W$ bosons,
\begin{eqnarray}
&& {t_p}\bar{t}_p \to bWth \to (bW)(bWbb) \to bbbbl\nu jj, \\
&& {t_p}\bar{t}_p \to thth \to (bWbb)(bWbb) \to bbbbbbl\nu jj.
\label{eq:3b-decaymode}
\end{eqnarray}
As in cases of single-b-jet and two-b-jet channels, we impose following selection-criteria with adopting the same method to reconstruct the invariant mass $M_{bl\nu}$,
\begin{enumerate}
\item There is only one lepton which is required to have $p_T >$ 30 GeV.
\item Transverse missing energy $\slashed{E}_T$ should be larger than 20 GeV.
\item There are more than four jets, and at least three of them are b-tagged.
\item The b-tagged jets are required to have $p_T >$ 140, 80, 80 GeV.
\item The invariant mass $M_{bl\nu}$ should be larger than 250 GeV.
\item The effective mass $M_{\rm eff}$ should be larger than 1000 GeV.
\end{enumerate}

\begin{figure}[p]
\begin{center}
\includegraphics[scale=0.37]{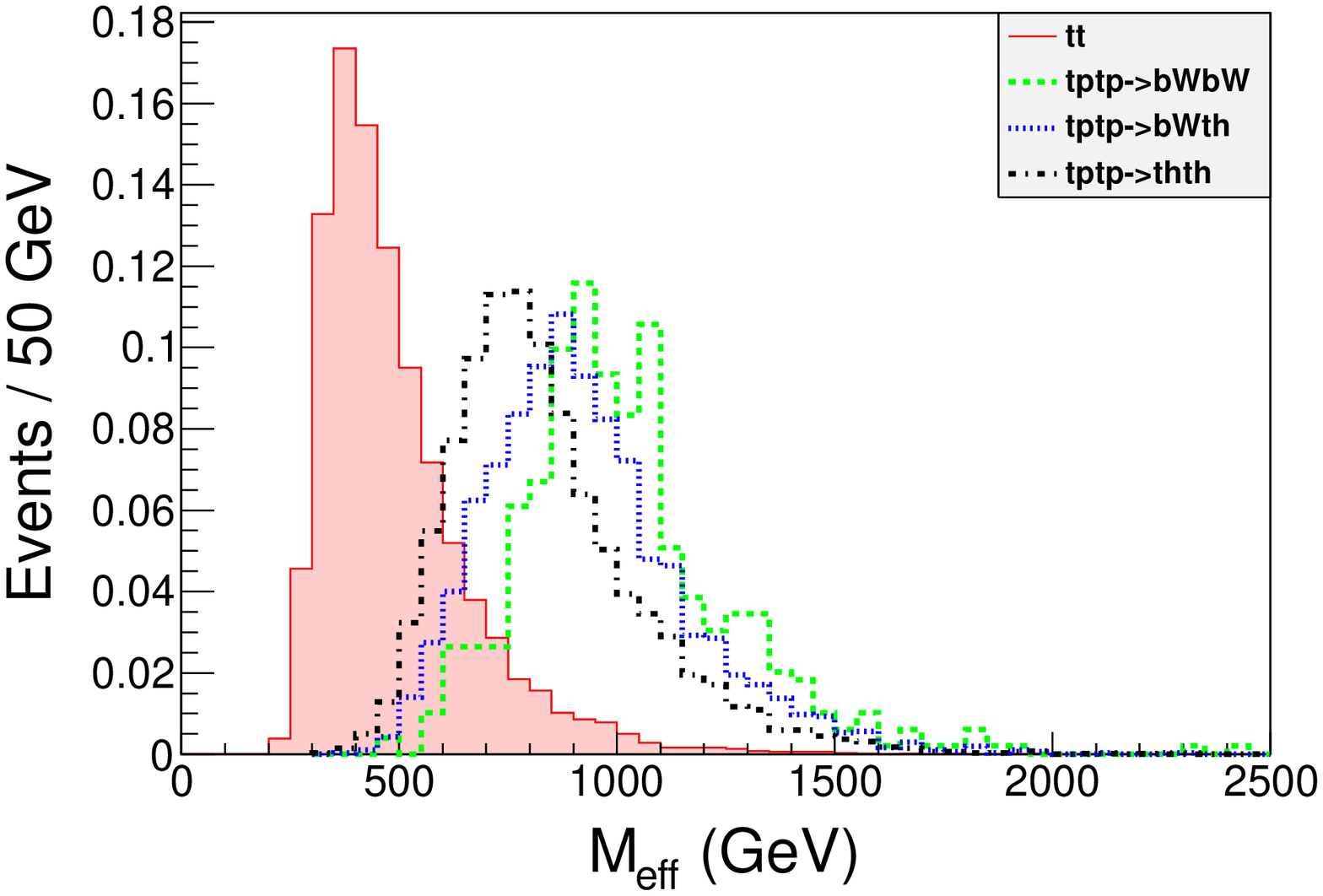}
\includegraphics[scale=0.37]{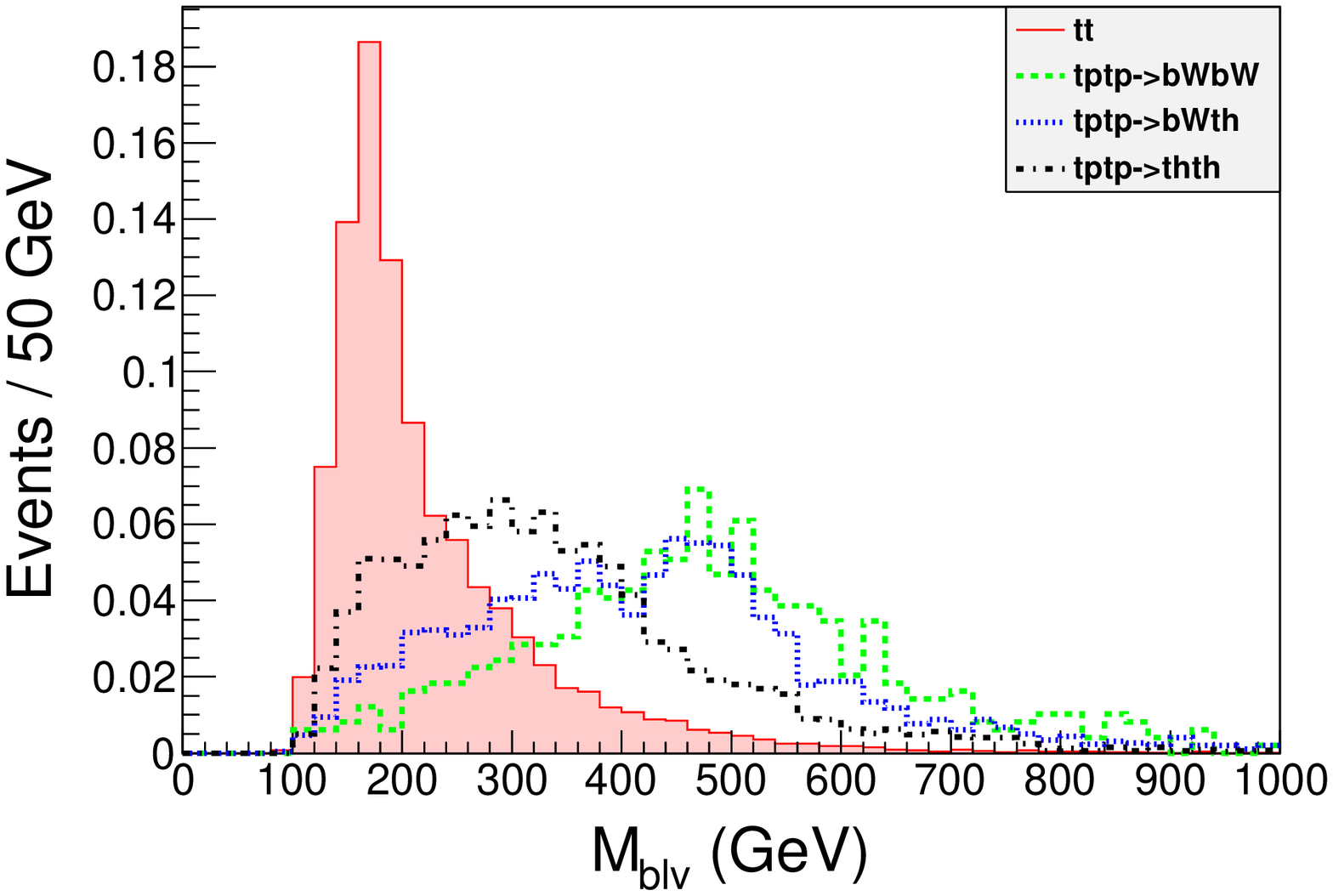} \\
\vspace{1.0cm}
\includegraphics[scale=0.37]{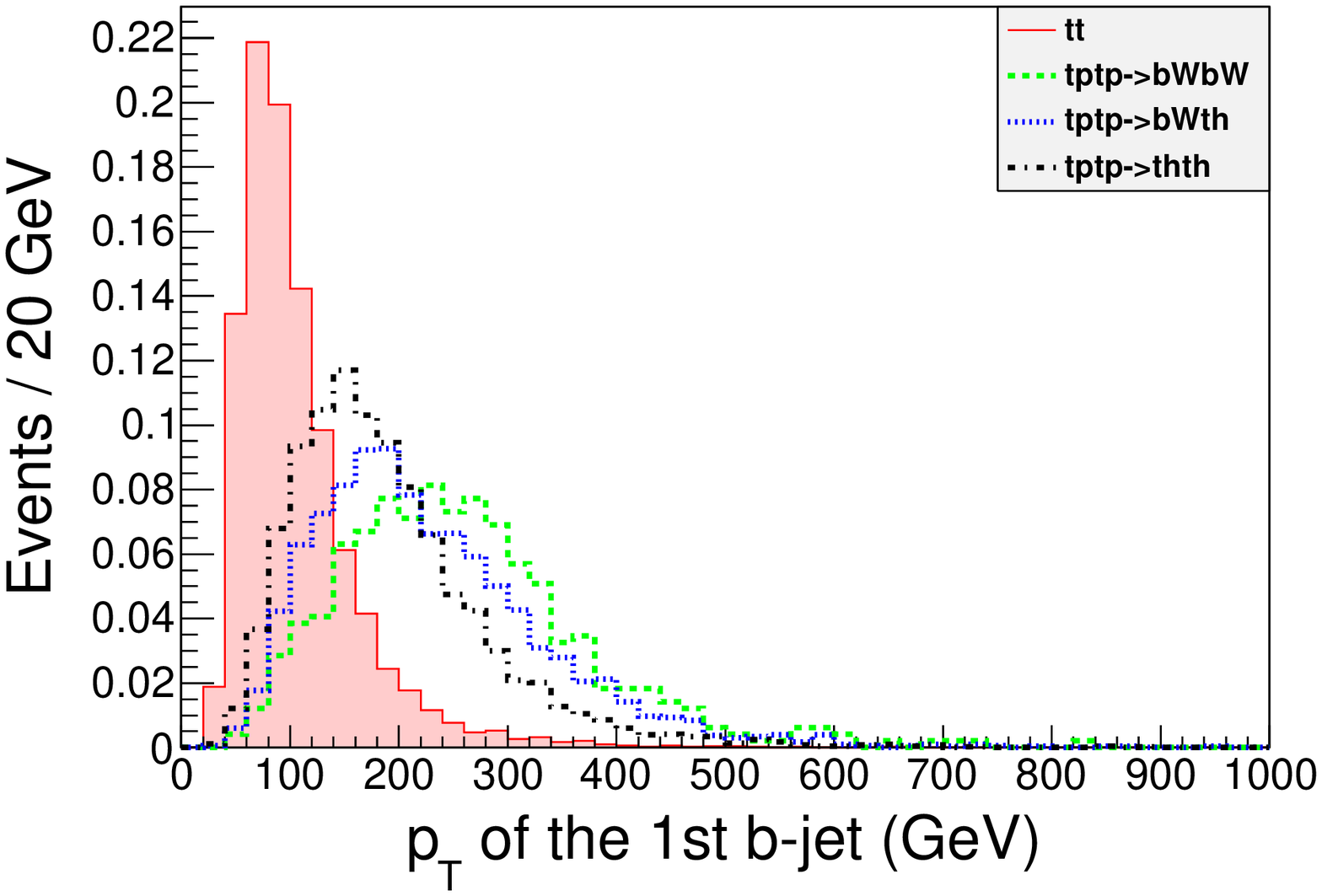}
\includegraphics[scale=0.37]{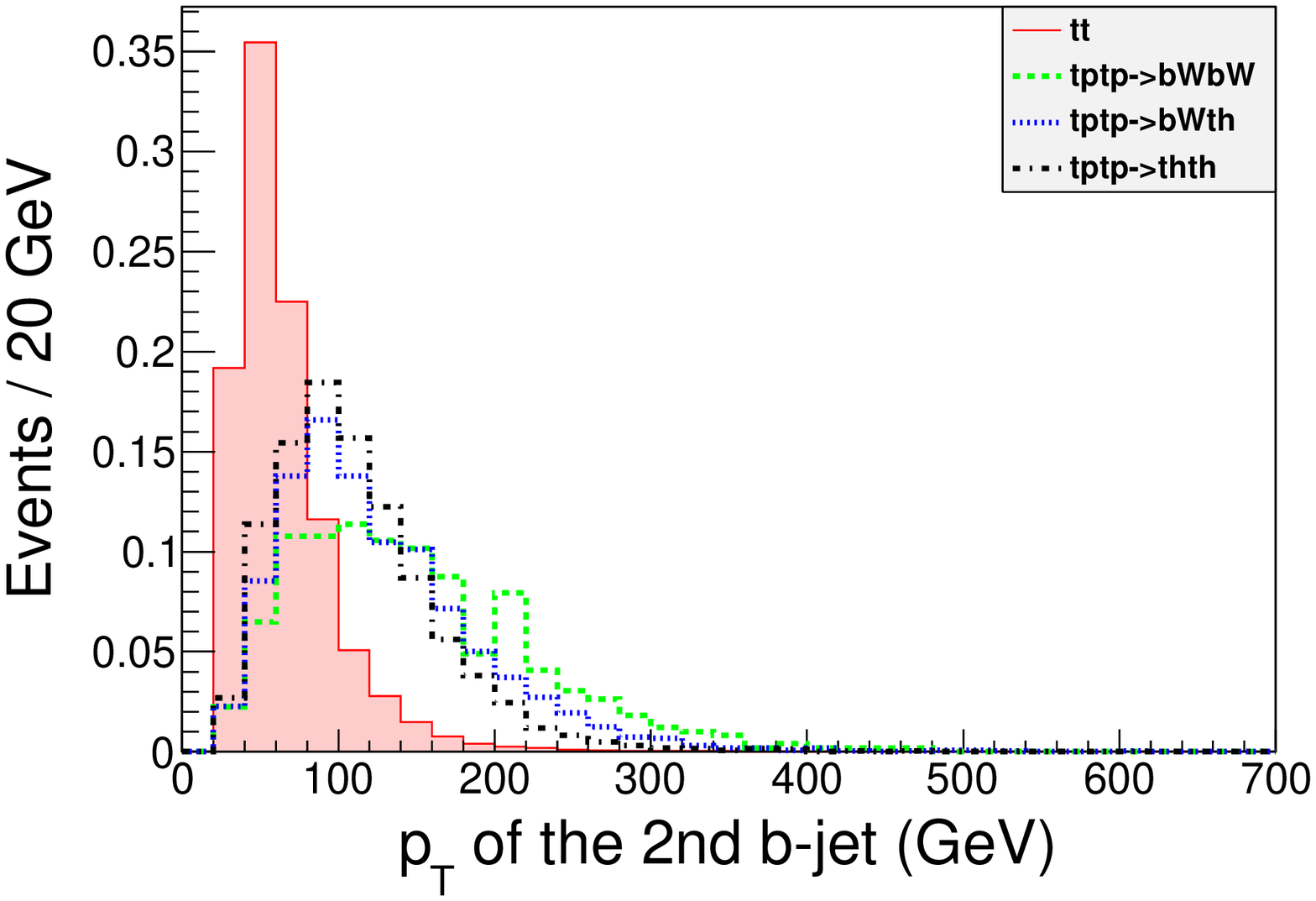} \\
\vspace{1.0cm}
\includegraphics[scale=0.37]{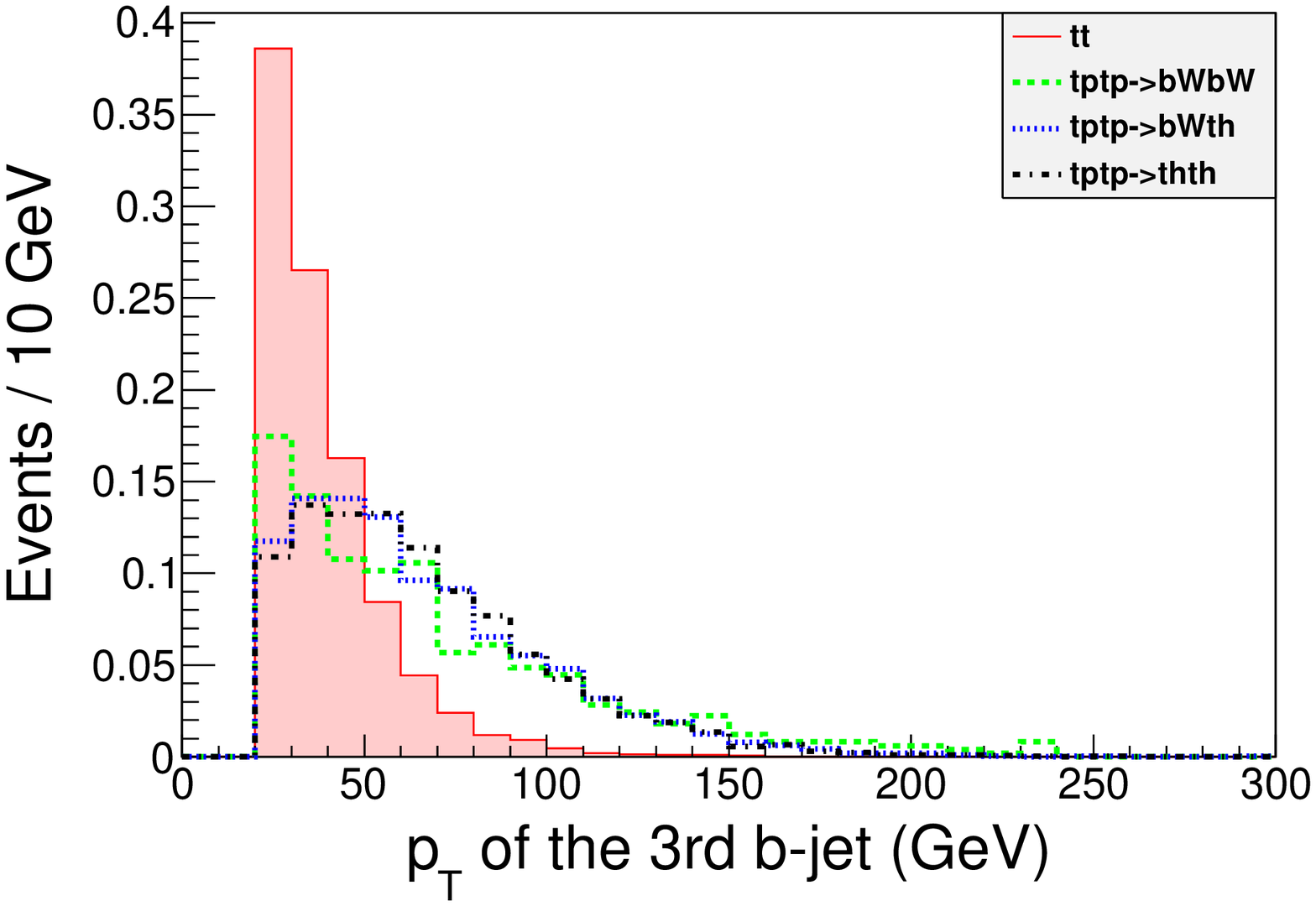}
\caption{\small
Distributions of $M_{\rm eff}$, $M_{bl\nu}$, and transverse momenta of the first to third b-jets for signal and background ($t\bar{t}+$jets) events after applying the cuts 1--3 in the analysis of the `1 lepton + $\geq$3b-jets' channel. The mass of the top partner is $m_{tp} =$ 500 GeV. Distributions of $t_p \bar{t}_p \to$ $bWbW$, $bWth$ and $thth$ are shown separately assuming all $t_p \bar{t}_p$-decays in each mode. The distributions are normalized so that their integrated values become unity.}
\label{fig: 3b-dist}
\end{center}
\end{figure}

Distributions of $M_{\rm eff}$, $M_{bl\nu}$, and transverse momenta of the first to third b-jets after applying kinematical cuts 1--3 are plotted in Fig.\ref{fig: 3b-dist}. As in the cases of single-b-jet and two-b-jet analyses, the mass of the top partner for signal events is $m_{t_p} =$ 500 GeV, and we concider decay patterns of $t_p\bar{t}_p\to$ $bWbW$, $bWth$, and $thth$.  The distribution of $t\bar{t}+$jets events is plotted in the same figure. The cut flows of the signal events and the main background events ($t\bar{t}+$jets and $t\bar{t}b\bar{b}$) are shown in Table \ref{table: 3b-cutflow}. Numbers of events for signals and backgrounds in the third low correspond to those with the integrated luminosity of 15 fb$^{-1}$, while their acceptances after applying all kinematical cuts 1--6 are shown in the last low. The acceptances in each decay pattern of $t_p\bar{t}_p$ are found in Table \ref{table: 3b-acceptance}.

\begin{table}[t]
\begin{center}
{\small
\begin{tabular}{c|ccccc}
& $t\bar{t}$ & $t\bar{t}b\bar{b}$ & $T\bar{T} \to bWbW$ & $T\bar{T} \to bWth$
& $T\bar{T} \to thth$ \\
\hline
Generated    & 6163292  & 90000   & 60000 & 60000 & 60000 \\
Without cuts & 3060000   & 24000   & 7650   & 7650   & 7650 \\
Cuts 1--3    & 5328      & 558     & 63     & 380    & 686 \\
Cut 4        & 123       & 26      & 19     & 99     & 170 \\
Cut 5        & 78        & 18      & 17     & 86     & 135 \\
Cut 6        & 15        & 4       & 11     & 40     & 45 \\
\hline
Acceptance   & 0.0000050 & 0.00016 & 0.0014 & 0.0052 & 0.0059 \\
\hline
\end{tabular}
}
\caption{\small Cut flows of signal and background events in the analysis of the `1 lepton + $\geq$3b-jets' channel. The mass of the top partner is $m_{tp} =$ 500 GeV. Results of $t_p \bar{t}_p \to$ $bWbW$, $bWth$ and $thth$ are shown separately assuming all $t_p \bar{t}_p$-decays in each mode.}
\label{table: 3b-cutflow}
\end{center}
\end{table}

\begin{table}[t]
\begin{center}
{\small
\begin{tabular}{c|cccccccc}
Mass (GeV) & 400 & 450 & 500 & 550 & 600 & 650 & 700 & 750 \\
\hline
$T\bar{T} \to bWbW$ & 0.00070 & 0.0014 & 0.0014 & 0.0022 & 0.0029 & 0.0034
& 0.0039 & 0.0042 \\
$T\bar{T} \to bWth$ & 0.0018 & 0.0035 & 0.0052 & 0.0098 & 0.013 & 0.016
& 0.019 & 0.022 \\
$T\bar{T} \to thth$ & 0.0014 & 0.0034 & 0.0059 & 0.0092 & 0.015 & 0.021
& 0.026 & 0.031 \\
\hline
\end{tabular}
}
\caption{\small Acceptances of signal events in the `1 lepton + $\geq $3b-jets' channel with several choices of the top partner mass. Results of $t_p \bar{t}_p \to$ $bWbW$, $bWth$ and $thth$ are shown separately assuming all $t_p \bar{t}_p$-decays in each mode.}
\label{table: 3b-acceptance}
\end{center}
\end{table}

\subsection{LHC-sensitivity to the top partner at 8 TeV}
\label{sec: prospects}

We are now at the position to discuss the LHC-sensitivity to the top partner.  In the following analysis, the systematic uncertainty of the SM backgrounds is simply set to be 20\% in all of these channels. This value is based on that obtained in the CMS experiment~\cite{CMS1} for the 1 lepton + $\geq$1b-jet channel.
Exclusion region is defined as the
one which is excluded at 95\% C.L. if the observed number of the events
is equal to the expected number of the SM background events.
Discovery region is the one where the expected number of the signal and
the SM background events deviates from the expected number of the SM
background events at 5$\sigma$ level.
For this calculation, we utilized the method reviewed in Ref.\cite{Choudalakis:2011bf}.

\begin{figure}[p]
\begin{center}
\includegraphics[scale=0.15]{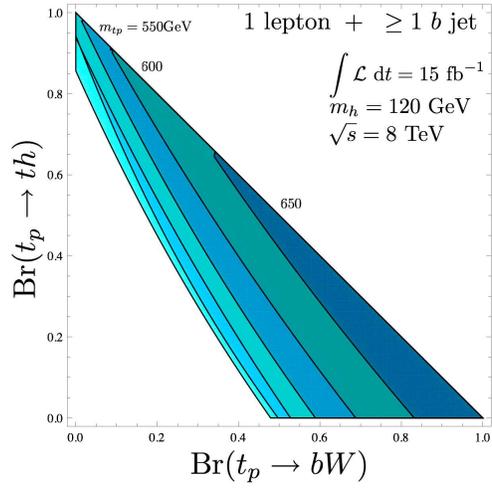} \\
\vspace{0.5cm}
\includegraphics[scale=0.15]{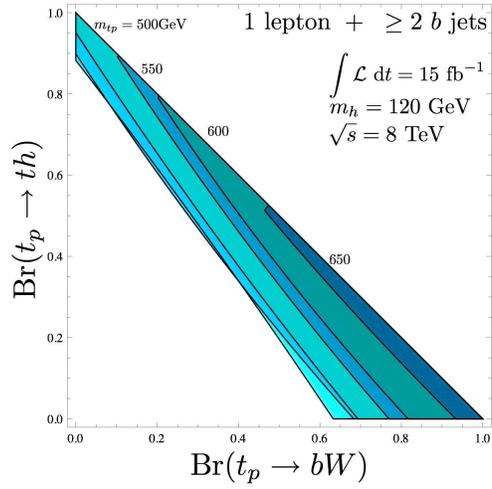} \\
\vspace{0.5cm}
\includegraphics[scale=0.15]{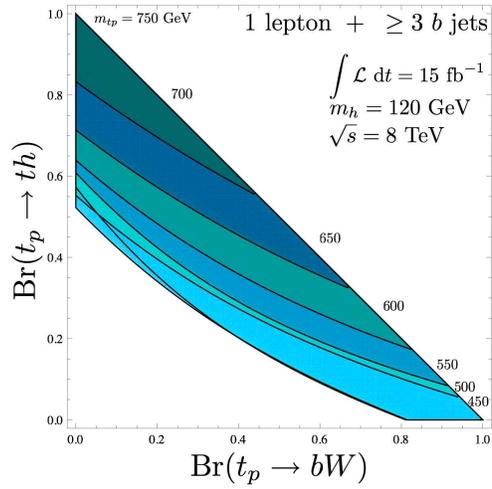}
\caption{\small Regions which would be excluded by the analysis (1 lepton + $\geq$1b-jet) and that of the multi-b-jet channels (1 lepton + $\geq$2 or 3b-jets) at 95\% C.L. with the integrated luminosity of 15 fb$^{-1}$. 
Contours are plotted in the region of Br($t_p \to bW$)+Br($t_p \to th$)$\leq$1.
The center of mass energy is 8 TeV.}
\label{fig: exclusion}
\end{center}
\end{figure}

Based on the SM backgrounds evaluated in the previous subsection,
expected exclusion-limits on signal cross sections in the single-, two-,
and three-b-jet channels at 95\% C.L. when no signal is detected turn out to be
\begin{eqnarray}
({\rm Cross \, section} \times {\rm Acceptance}) \, >
\left\{
\begin{array}{rl}
4.4 \, {\rm fb} :  & {\rm 1~lepton \, + \, \geq 1b\mathchar`-jet} \\
1.2 \, {\rm fb} :  & {\rm 1~lepton \, + \, \geq 2b\mathchar`-jets} \\
0.92 \, {\rm fb} : & {\rm 1~lepton \, + \, \geq 3b\mathchar`-jets}.
\end{array}
\right.
\end{eqnarray}
For the integrate luminosity of 15 fb$^{-1}$, the regions which would be
constrained by the analysis of the single-b-jet channel (1 lepton + $\geq$1b-jet), two-b-jet channels (1 lepton + $\geq$2b-jets), and three-b-jet channels (1 lepton + $\geq$3b-jets) are plotted in upper, middle, and lower panels of Fig.\ref{fig: exclusion}, respectively. The regions are depicted on the plane of Br($t_p \to bW$) and Br($t_p \to th$) for various values of the top partner mass. 
The single- and two-b-jet channels give stringent constraints on $m_{tp}$ when the top partner decays dominantly into $bW$, while the three-b-jet channels constrains the top partner when Br(${t_p} \to th$) is large. The LHC experiment has a capability to exclude $m_{tp}$ less than 750 (650) GeV when ${t_p} \to th (bW)$ is dominant. 

In this study, we use a fast detector simulator, Delphes, and therefore our result should be regarded as order estimate. Especially the discovery potential of the top partner dominantly decaying into $th$ depends on b-tagging/mis-tagging efficiencies. They must be experimentally determined.  Some of our estimate is fragile. For example, in the lower panel, when the top partner is lighter than 500 GeV, some regions are still constrained even if the decay mode ${t_p} \to th$ is forbidden.  These constraints are due to the mis-tagging of light-jets, which are regarded as b-jets, and are expected to suffer from large uncertainty of our Monte-Carlo simulations. However, these regions are covered by analyses of single- and two-b-jet channels and the ambiguities do not affect the search of the top partner.

\begin{figure}[t]
\begin{center}
\includegraphics[scale=0.15]{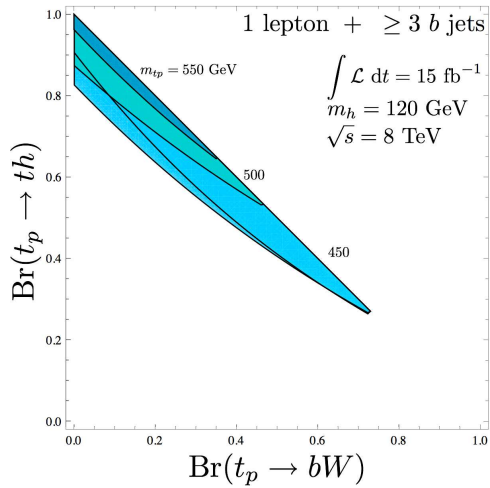}
\caption{\small The region in which the signal could be discovered in the three-b-jet channels (1 lepton + $\geq$3b-jets) at the 5$\sigma$ level with the integrated luminosity of 15 fb$^{-1}$.}
\label{fig: discovery}
\end{center}
\end{figure}

We also consider the capability of the LHC experiment to discover the signal with the center of mass energy of 8 TeV. In the single-b-jet and multi-b-jet channels, signal events can deviate from those of the SM backgrounds at 5$\sigma$ level when their corresponding cross sections (times acceptances) satisfy following conditions,
\begin{eqnarray}
({\rm Cross \, section} \times {\rm Acceptance}) >
\left\{
\begin{array}{rl}
36 \, {\rm fb}  : & {\rm 1~lepton \, + \, \geq 1b\mathchar`-jet} \\
3.6 \, {\rm fb} : & {\rm 1~lepton \, + \, \geq 2b\mathchar`-jets} \\
2.3 \, {\rm fb} : & {\rm 1~lepton \, + \, \geq 3b\mathchar`-jets}.
\end{array}
\right.
\end{eqnarray}
The region in which the top partner signal could be discovered at 5$\sigma$ level with the use of the three-b-jet channels is plotted in Fig.\ref{fig: discovery} for various values of $m_{tp}$. No region for the discovery of the top partner with the mass greater than 400 GeV can be found in the analyses of the single- and two-b-jet channels. With the integrated luminosity of 15 fb$^{-1}$ at 8 TeV, the LHC experiment has the capability to  the signal up to $m_{tp} <$ 550 GeV when the decay mode ${t_p} \to th$ dominants.

\section{Implications to new physics models}
\label{sec: application}

We discuss some implications of the results obtained in the previous
 section to new physics models beyond the SM. Based on the effective
 action of the top partner given in Section 2, we clarify parameter
 regions which can be covered by the LHC at 8 TeV with 15 fb$^{-1}$ data.

\begin{figure}[t]
\begin{center}
\includegraphics[scale=0.16]{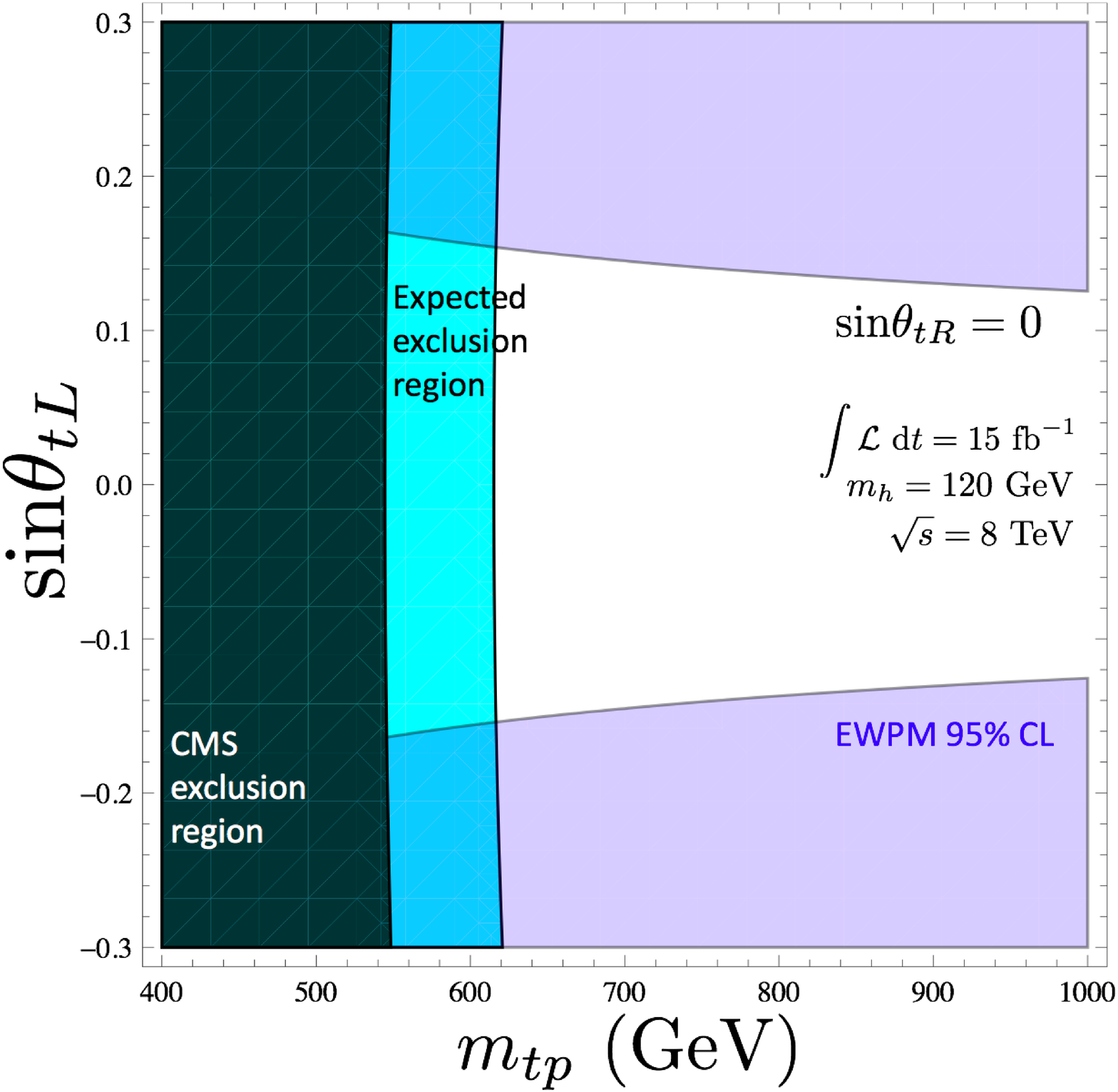}
\qquad
\includegraphics[scale=0.16]{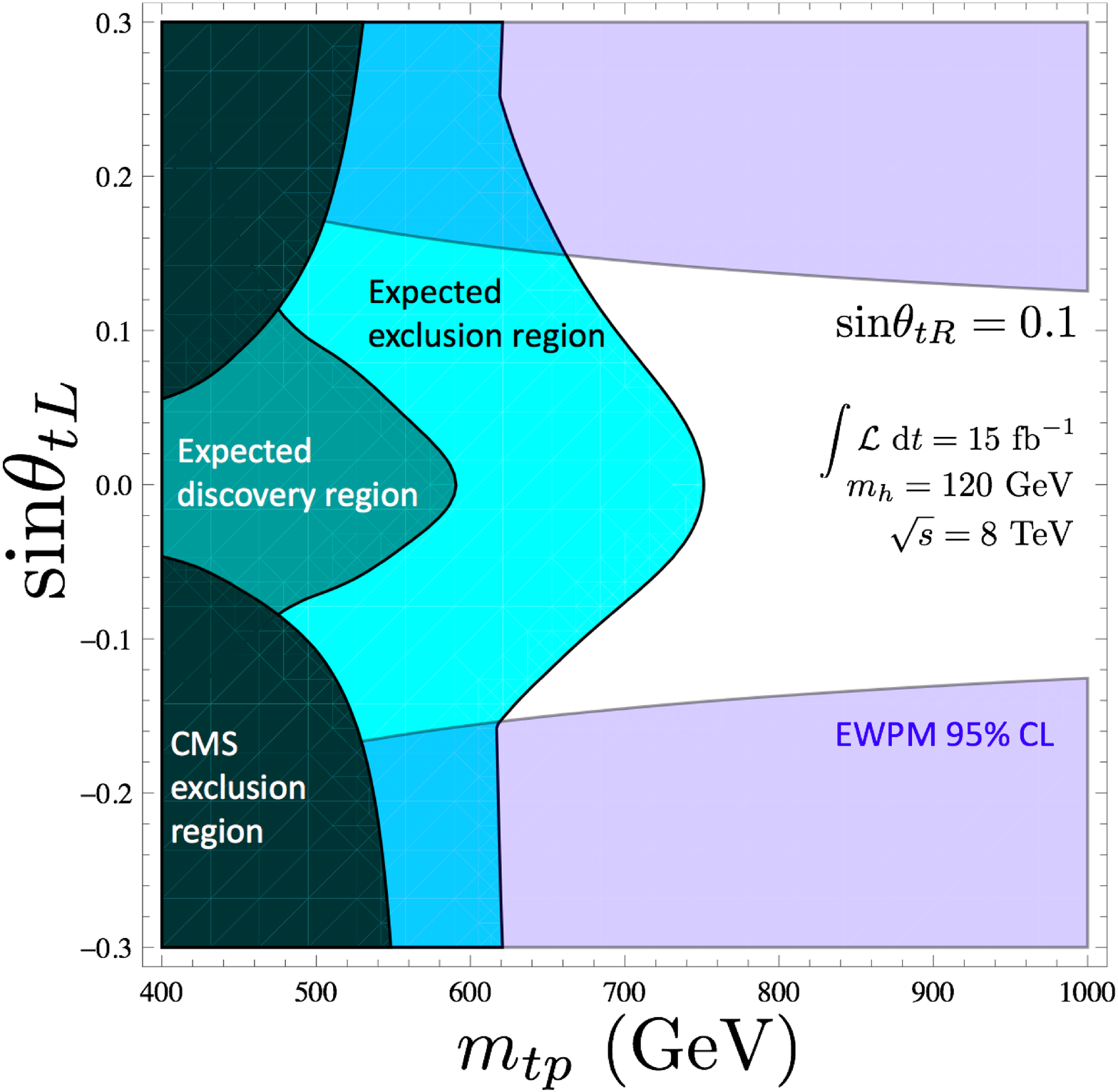}
\caption{\small Expected exclusion region at 95\% C.L. and discovery  region at 5$\sigma$ level. The angle $\sin \theta_{tR}$ is 0 (0.1) in the left (right) panel.}
\label{fig: exclusion-app}
\end{center}
\end{figure}

The expected exclusion region at 95\% C.L. and the discovery  region at 5$\sigma$ level are shown in Fig.\ref{fig: exclusion-app} on the ($m_{tp}$, $\sin \theta_{tL}$)-plane, which are obtained from analyses of all the single-, two-, and three-b-jet channels. As explainded in Section 2 and as seen in Fig.\ref{fig: branching ratio}, the rise of the right-handed mixing angle $\sin \theta_{tR}$ which strongly depends on the coefficient of the higher dimensional operator $\lambda/\Lambda$ significantly enhances the branching fraction of $t_p\to th$. The  $\sin \theta_{tR}$ is fixed to be 0 (0.1) in the left (right) panel of the figure. Parameter regions which have already been constrained by the electroweak precision measurement~\cite{Harigaya:2011yg} and the CMS experiment~\cite{CMS2} are also depicted in the panels. These figures show the LHC experiment will cover the top partner mass up to about 620 (750) GeV when $\sin \theta_{tR} =$ 0 (0.1).  In the little higgs model, the top partner mass larger than 600 GeV requires the fine tuning severer than 10\% \cite{Harigaya:2011yg, Big-corrections}.  
Also, we show the discovery region in the right panel when the mixing
angle $\sin \theta_{tR} =$0.1 .
This indicates that the multi-b-jet channels have potential to search for the top partner at the LHC.

\section{Summary}
\label{sec: summary}

We have investigated the capability of the LHC experiment to search for
the top partner with the center of mass energy  8 TeV and the integrated
luminosity of 15 fb$^{-1}$. In this paper we assume that the top partner
is the vector-like quark which is singlet (triplet) under the SM
SU(2)$_L$ (SU(3)$_c$) gauge group and is interacting only with the third
generation quarks. In order for the discussion to be as general as
possible, we have used the effective action to describe the physics of
the top partner. If new physics comes from the strong dynamics, the
dimension-five operator in Eq.(1) becomes sizable, which enhances the
decay mode of $t_p\to th$.
The $t_p\rightarrow th$ decay produces multiple b-jets in the final state.
We have found that the multi-b-jet channels play a complementary role to
the existing searches in the one-b-jet + one lepton channel and
multi-jet + three leptons channel which rely on the decay modes $t_p \to bW$ and $t_p \to tZ$. 

It also has been shown that the multi-b-jet channels are so powerful
that it is even possible to discover the top partner signal at 5$\sigma$
level when the top partner lighter than about 550 GeV dominantly decays
into $th$. With the use of results obtained in these analyses, we have
also discussed some implications to new physics models.  Using all the
single-, two-, and three-b-jet channels as summarized in Fig.\ref{fig:
exclusion-app}, we have found that wide regions of the model parameter
space will be excluded at 95\% C.L. if no signal is detected. The mass of top partner is, in fact, covered up to about 750 GeV for the large the right-handed mixing angle $\sin \theta_{tR}$ while up to about 620 GeV for the small $\sin \theta_{tR}$.

\section*{Acknowledgement}

We thank Shoji Asai, Hitoshi Murayama, Satoshi Shirai, Satoru Yamashita and Tsutomu T. Yanagida for valuable discussions. This work is supported by Grant-in-Aid for Scientific research from the Ministry of Education, Science, Sports, and Culture (MEXT), Japan (Nos. 23740169, 22244021 for S.M. and Nos. 22540300, 23104006 for M.M.N.), and also by World Premier International Research Center Initiative (WPI Initiative), MEXT, Japan. The work of K.H. and K.T. is  supported in part by JSPS Research Fellowships for Young Scientists.

\appendix

\section{Validating our simulation framework}
\label{app: validation}

In order to validate our simulation framework, we present the
the transverse mass distributions of the SM processes. These distributions are compared to those
provided by the ATLAS collaboration in Ref.\cite{ATLAS-validation}, with
particularly focusing on events involving missing transverse momentum,
b-jets, and a lepton.
The validation with this channel is important since we also treat the
same objects in our analysis given in Sec.~\ref{sec: simulation}.
Selection-criteria, which are imposed in order to compare with the ATLAS result, are given by
\begin{enumerate}
\item There is only one electron (muon) which should have $p_T >$ 25 (20) GeV.
\item There are at least four jets which are required to have $p_T >$ 50 GeV.
\item At least one of the jets is b-tagged with $p_T >$ 50 GeV and $|\eta| >$ 2.5.
\item Transverse missing energy $\slashed{E}_T$ should be larger than 80 GeV.
\item The transverse mass $m_T$ should be larger than 100 GeV.
\item The effective mass $M_{\rm eff}$ should be larger than 600 GeV.
\end{enumerate} 
The effective mass $M_{\rm eff}$ is defined by the scalar-sum of the
missing energy, the transverse momenta of leading four jets, and that of
the lepton, while the transverse mass $m_T$ is calculated from the
transverse missing momentum and the lepton momentum.
In the analysis conducted by the ATLAS collaboration, jets pointing to the region where the
electromagnetic calorimeter does not work properly
are removed. The loss of the acceptance caused by this selection is as
large as 20\%. We take this effect into account just by multiplying the number of events after the selections by 0.8.

\begin{figure}[t]
\begin{center}
\includegraphics[scale=0.37]{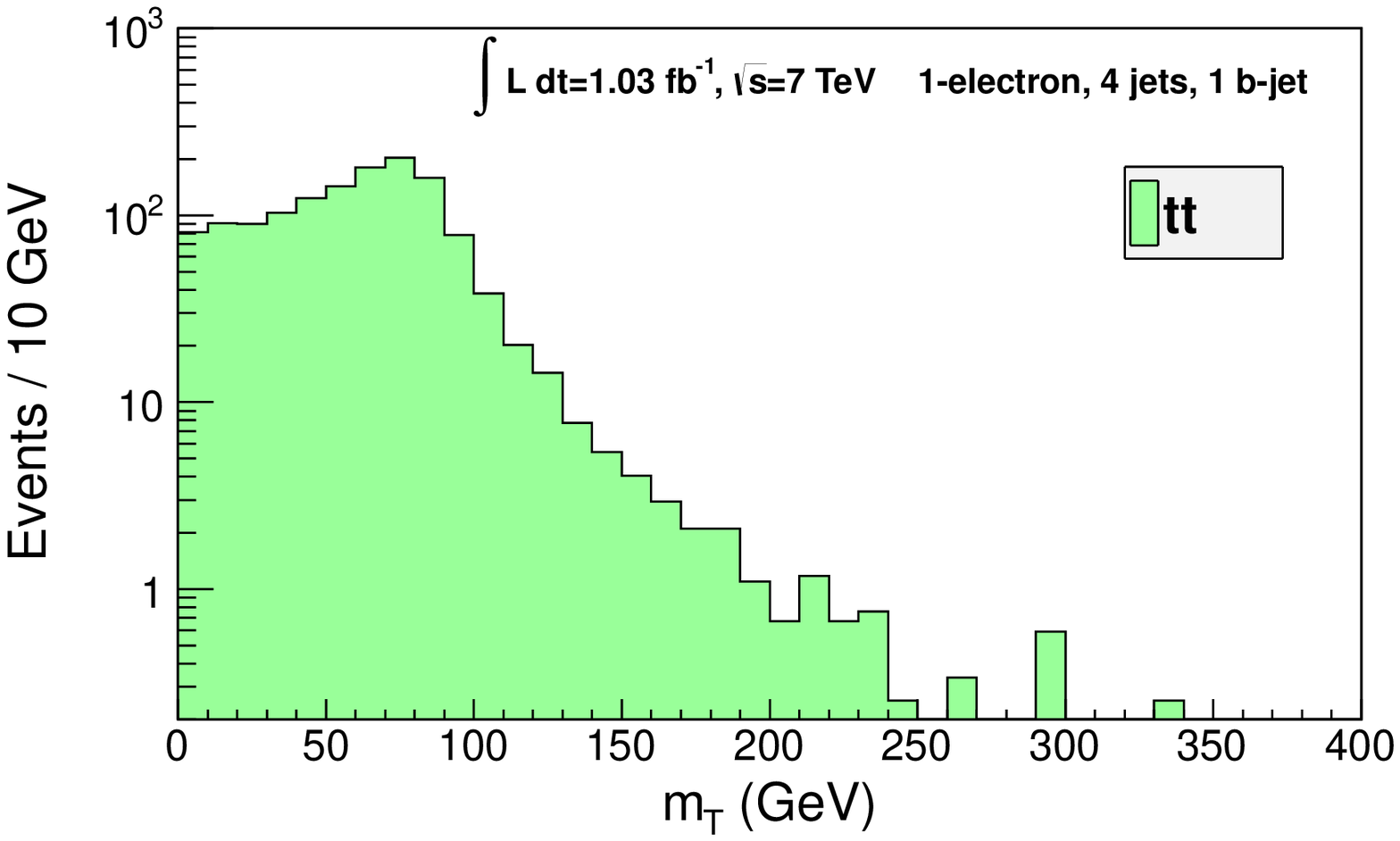}
\includegraphics[scale=0.37]{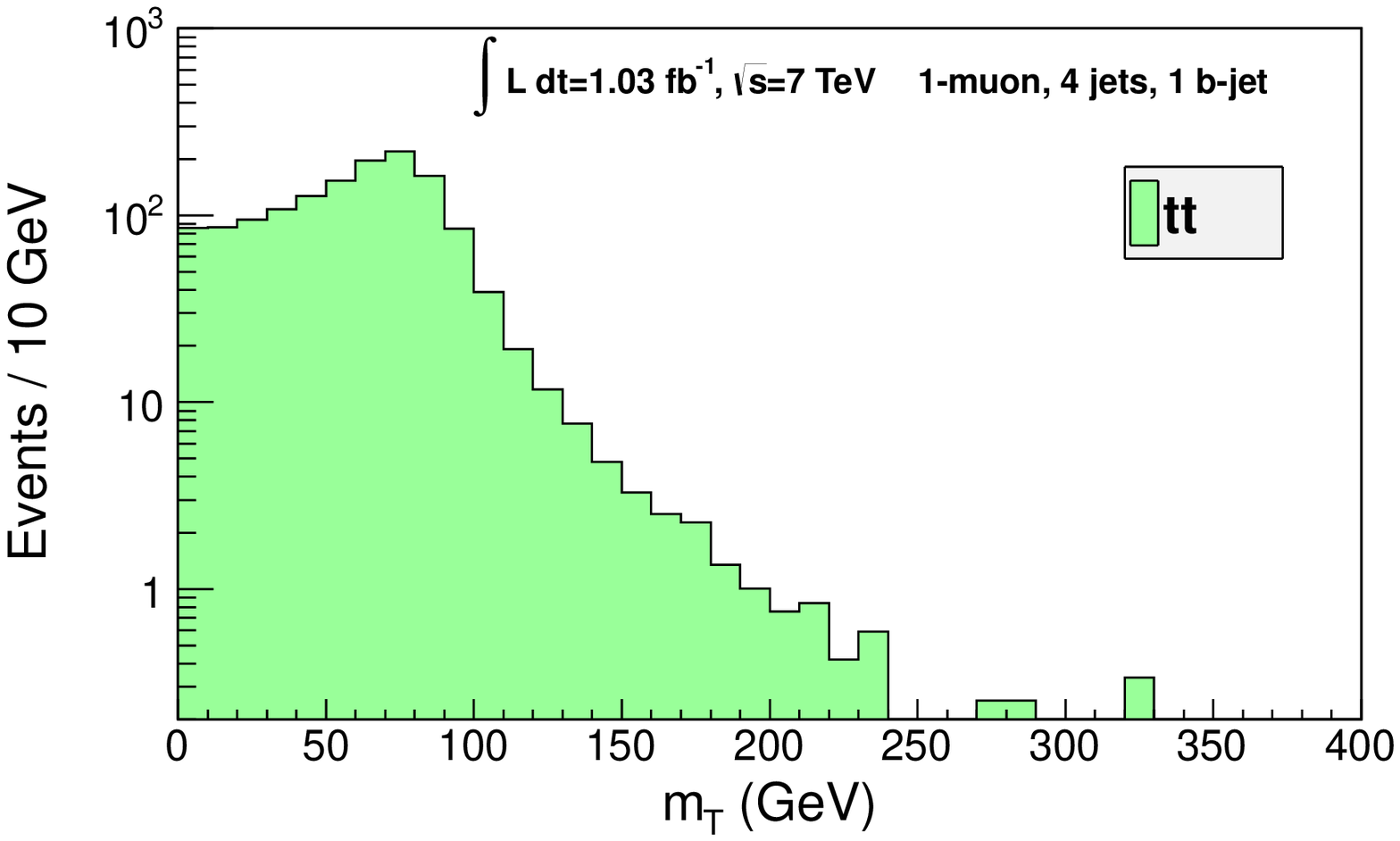} \\
\includegraphics[scale=0.37]{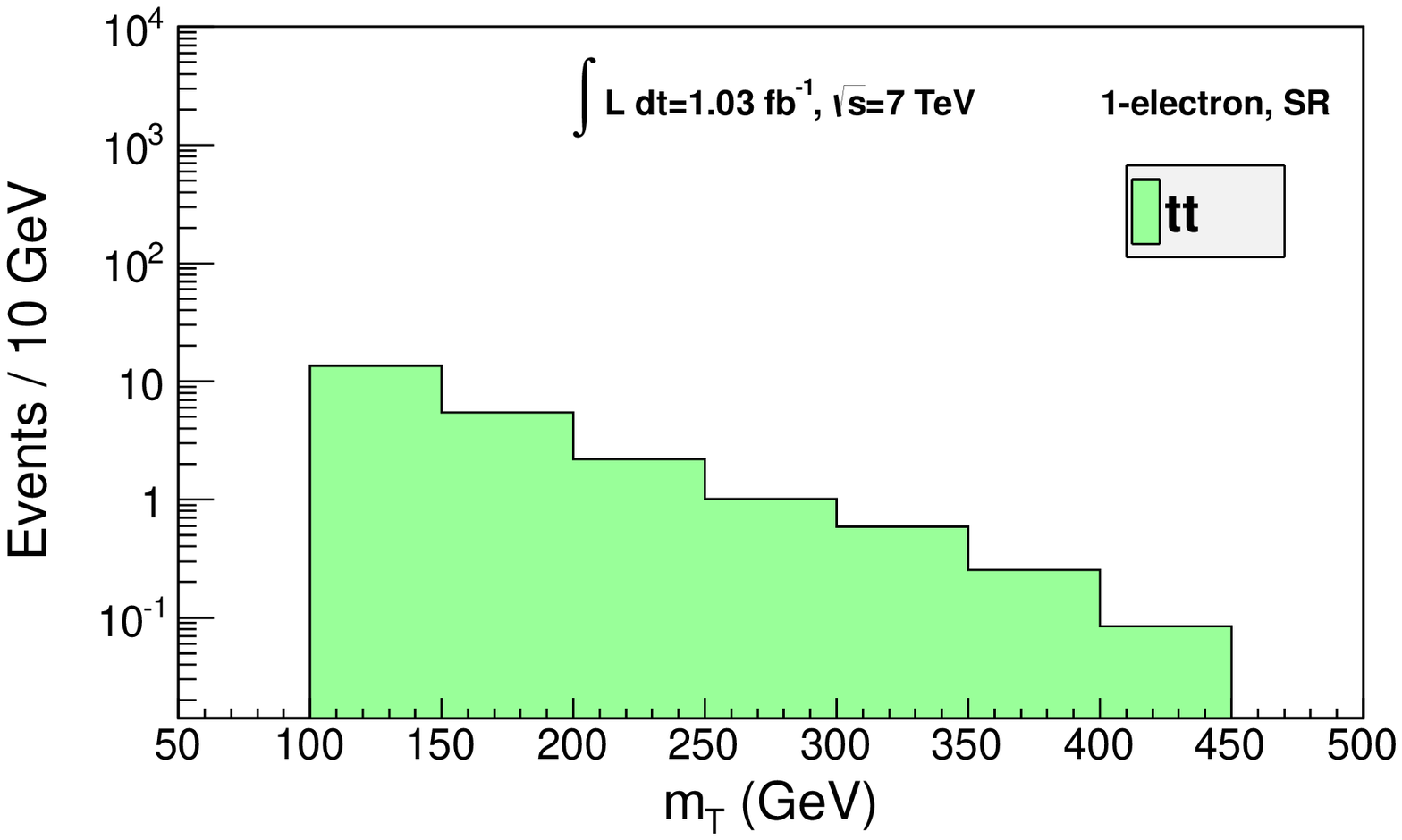}
\includegraphics[scale=0.37]{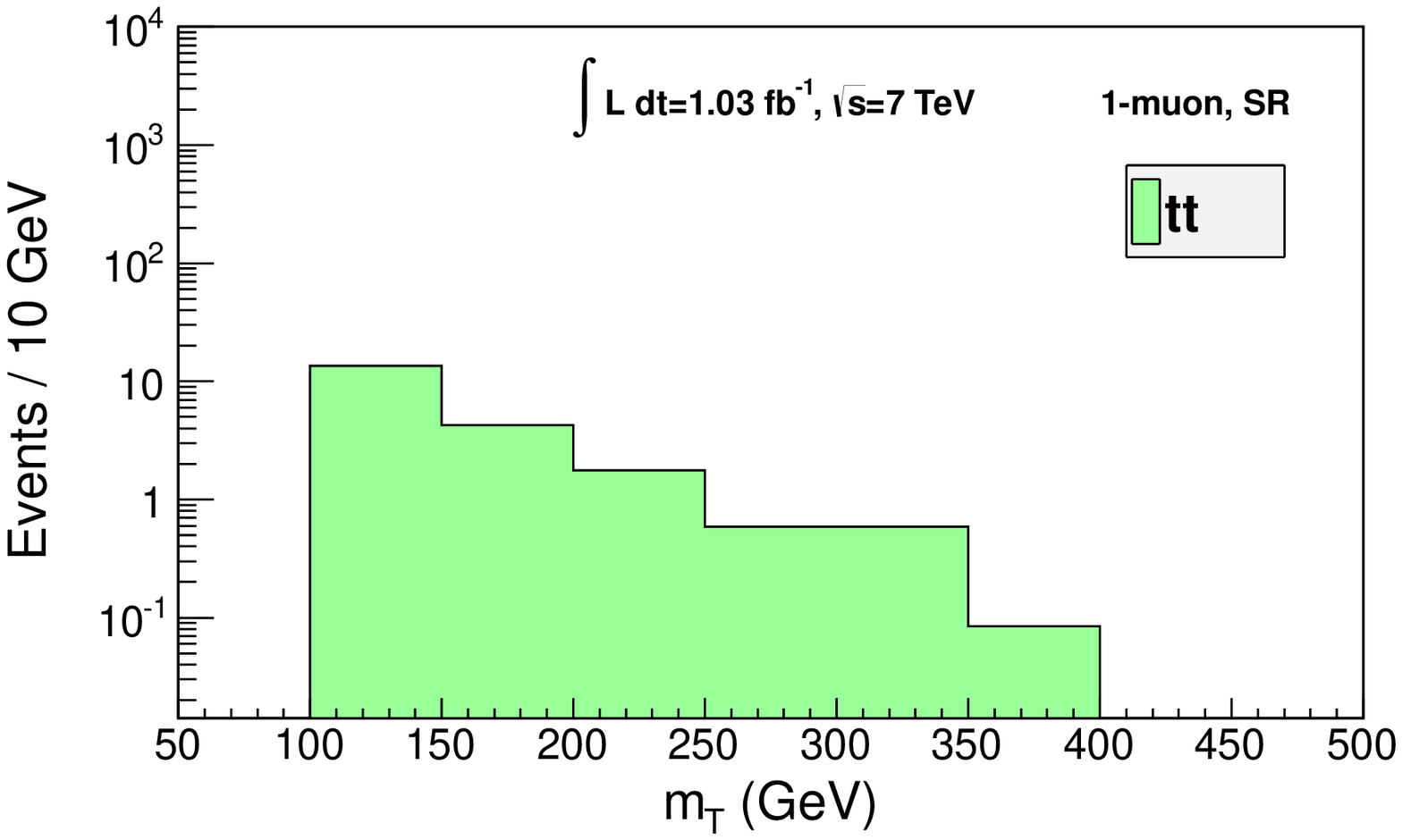}
\caption{\small Distributions of the transverse mass after imposing
 kinematical cuts 1--3 (upper two panels) and cuts 1--6 (lower two
 panels). The left two panels show distributions involving an electron,
 while the right two panels show those involving a muon.}
\label{fig: mT}
\end{center}
\end{figure}

The distributions of the transverse mass after imposing kinematical cuts
1--3 and after cuts 1--6 are plotted in upper two panels and lower two
panels of Fig.\ref{fig: mT}, respectively. The distributions of
$t\bar{t}+$jets events are depicted in the
figures. The left two plots show distributions involving an electron, while
right two plots show those with a muon. These figures should be compared
to Fig. 3 and 5 in Ref.~\cite{ATLAS-validation}. Our results are consistent with those given by the ATLAS collaboration within the accuracy of 10--20\%.

\section{CMS-based analysis}
\label{app: CMS-based analysis}

In this appendix, we show results based on the analysis in the `1 lepton
+ $\geq$1 b-jet' channel using the kinematical variables $M_{{\rm fit}}$
introduced in the
CMS-analysis. We impose a following selection-criteria to generated
events subsequantly,
\begin{enumerate}
\item There is only one lepton which is required to have $p_T >$ 30 GeV.
\item Transverse missing energy $\slashed{E}_T$ should be larger than 20 GeV.
\item There are more than three jets, and at least one of them are b-tagged.
\item Leading four jets are required to have $p_T >$ 80, 50, 30, and 30 GeV.
\item The leading b-jet is required to have $p_T >$ 300 GeV.
\item The kinematical variable $M_{\rm fit}$ should be larger than 350 GeV
\item The effective mass $M_{\rm eff}$ should be larger than 1300 GeV.
\end{enumerate}
Here, the effective mass is defined by $M_{\rm eff} =
p_{T\mathchar`-{\rm lepton}} + \slashed{E}_T + \sum_{i = 1}^{{\rm
min}(5,N_{{\rm jets}})} p_{T\mathchar`-i{\rm th \, j}}$. The kinematical variable $M_{\rm fit}$ is defined as $M_{\rm
fit} = \min (m_{l\nu b},m_{jjb})$, where the two invariant masses are calculated
from all possible combination of jets which minimize following chi-2 function,
\begin{eqnarray}
\chi^2 =
\left( \frac{m_{lv} - m_W}{\Delta m_{lv}} \right)^2
+ \left( \frac{m_{jj} - m_W}{\Delta m_{jj}} \right)^2
+ \left(
\frac{m_{l\nu b} - m_{jjb}}{[(\Delta m_{l\nu b})^2 + (\Delta m_{jjb})^2]^{1/2}}
\right)^2,
\end{eqnarray}
where uncertainties of the invariant mass $\Delta m_{ij}$ ($\Delta
m_{ijk}$) are evaluated by considering the resolutions of hadronic and
electromagnetic calorimeters of the ATLAS detector. The transverse
momentum of the neutrino is identified with the missing transverse
momentum, while the longitudinal one is treated as a free parameter
which is determined by minimizing the $\chi^2$-function. When only one
b-jet is detected, one of the other jets is regarded as another b-jet and is used to evaluate the above $\chi^2$-function.

\begin{figure}[t]
\begin{center}
\includegraphics[scale=0.43]{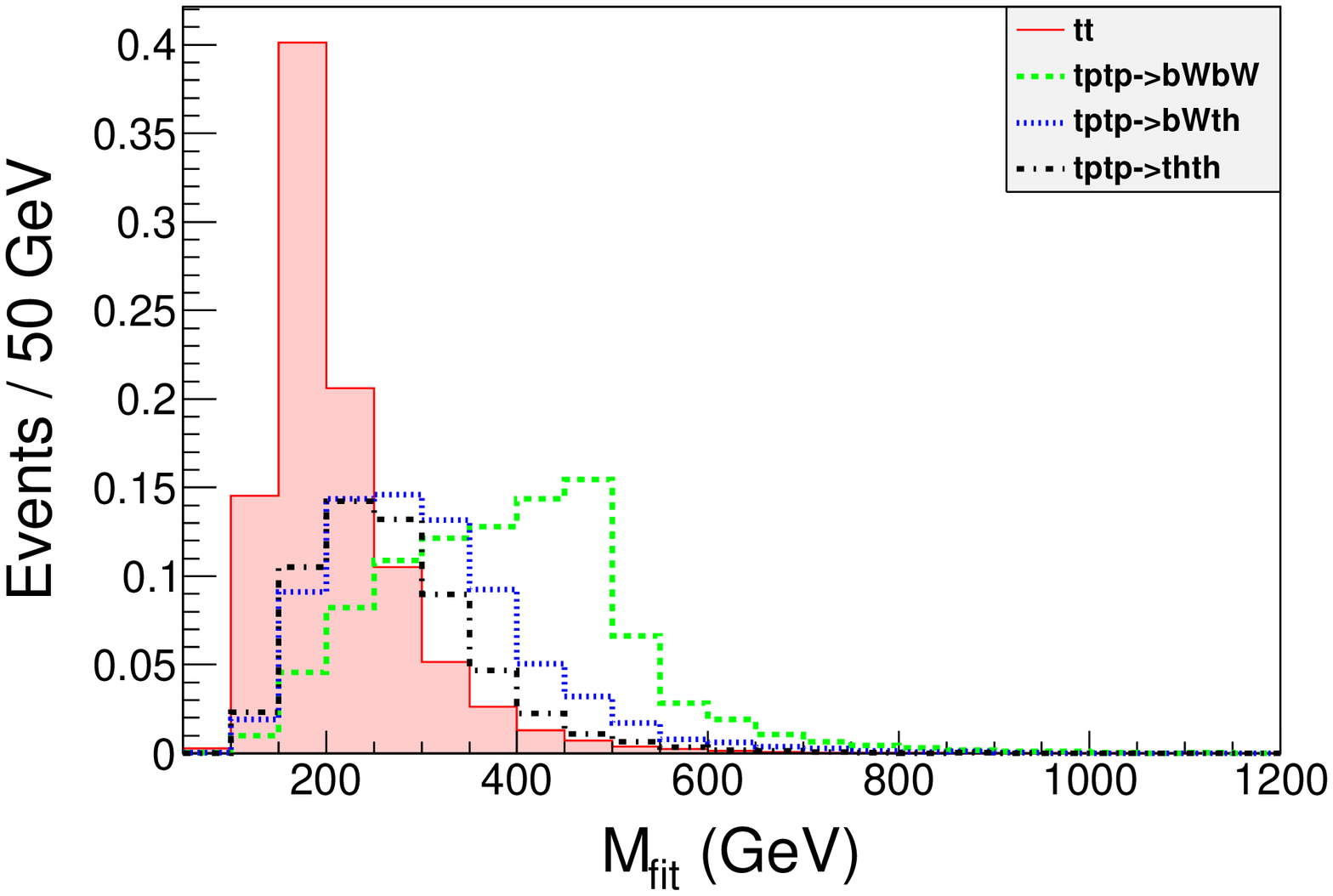}
\includegraphics[scale=0.14]{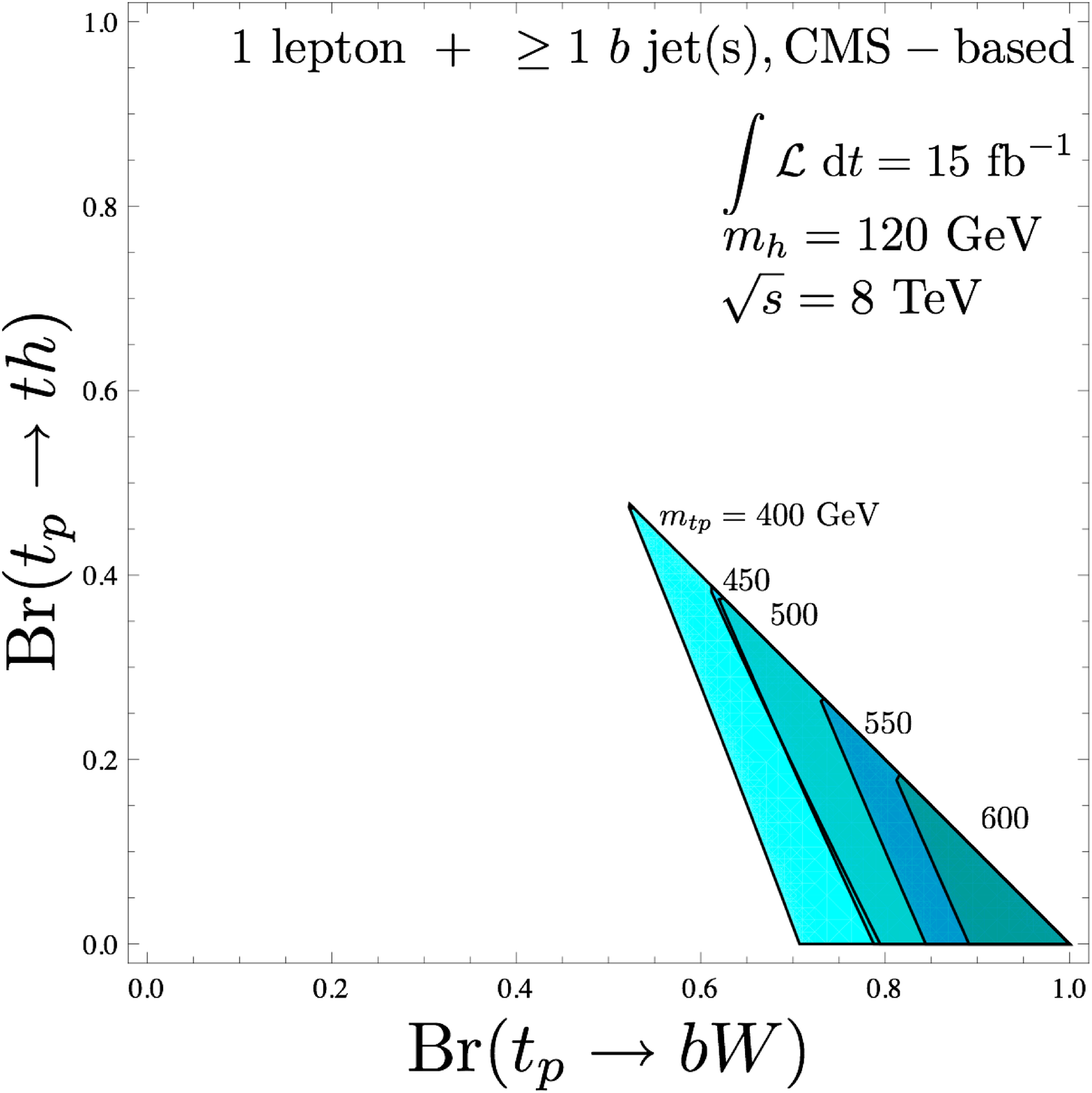}
\caption{\small ({\bf Left panel}) Distribution of $M_{\rm fit}$ for signal and background ($t\bar{t}+$jets) events after applying the cuts 1--4 in the `1 lepton
+ $\geq$1 b-jet' channel adopted in the CMS experiment. 
The mass of the top partner is  $m_{tp} =$ 500 GeV. Results of $t_p \bar{t}_p \to$ $bWbW$, $bWth$ and $thth$ are shown separately assuming all $t_p \bar{t}_p$-decays in each mode. The distribution is normalized so that its integrated value becomes unity. ({\bf Right panel}) Region which would be excluded by the conventional analysis with the integrated luminosity of 15 fb$^{-1}$.}
\label{fig: CMS-based}
\end{center}
\end{figure}

The distribution of $M_{\rm fit}$ after applying kinematical cuts 1--4
is plotted in the left panel of Fig.\ref{fig: CMS-based} for
$m_{tp}=500$ GeV. Other distributions such as $p_{T\mathchar`-1{\rm st \, b}}$ and $M_{\rm eff}$ are exactly the same as those in Fig.\ref{fig: 1b-dist}. The $M_{\rm fit}$distribution of $t\bar{t}+$jets events is also plotted in the same figure. On the other hand, in Table \ref{table: CMS-cutflow}, the cut flows of both signal and background events are shown. Numbers of events for the signal and background in the second low correspond to those assuming the integrated luminosity of 15 fb$^{-1}$, and their acceptances after applying all kinematical cuts are shown in the last low. Acceptances of the signal events for various masses of the top partner  in each decay pattern of ${t_p}\bar{t}_p$  are also found in Table~\ref{table: CMS-acceptance}.

Using this CMS-based analysis and assuming the systematic uncertainty of 20\%, the exclusion-limit on the production cross section of ${t_p}\bar{t}_p$ is given by
\begin{eqnarray}
({\rm Cross \, section} \times {\rm Acceptance}) \, > \,
2.4 \, {\rm fb} : {\rm 1~lepton \, + \, \geq 1b\mathchar`-jet} \, {\rm (CMS\mathchar`-based)}.
\end{eqnarray}
 The regions which would be constrained by the conventional analysis is
 plotted in the right panel of Fig.\ref{fig: CMS-based} with the integrate luminosity of 15 fb$^{-1}$. The region is depicted on the plane of Br(${t_p} \to bW$) and Br(${t_p} \to th$) for various values of the top partner mass. It can be seen that the conventional analysis gives milder constraints than those discussed in Sec. \ref{sec: simulation}. On the other hand, the condition that signal events can deviate from those of SM backgrounds at 5$\sigma$ level is given by
\begin{eqnarray}
({\rm Cross \, section} \times {\rm Acceptance}) \, > \,
11 \, {\rm fb} : {\rm 1~lepton \, + \, \geq 1b\mathchar`-jet} \, {\rm (CMS\mathchar`-based)}.
\end{eqnarray}

Note that the analysis given here is not exactly the same one as the
CMS one. In the CMS-analysis, high-$p_T$ cut on the b-jet (cut 5) is not
imposed. Furthermore, the CMS collaboration puts the bound on the cross
section of the signal by examining the distributions of $M_{{\rm fit}}$
and $M_{\rm eff}$, not by the number counting.
However, the analysis given here is sufficient in order to compare the sensitivity of the
analysis using $M_{\rm fit}$ with the one using $M_{{\rm bl\nu}}$.

\begin{table}[t]
\begin{center}
{\small
\begin{tabular}{c|ccccc}
& $t\bar{t}$ & $W$+jets& $T\bar{T} \to bWbW$ & $T\bar{T} \to bWth$
& $T\bar{T} \to thth$ \\
\hline
Without cuts & 3060000 & 4800000   & 7650   & 7650   & 7650 \\
Cuts 1--3    & 282861  & 16756     & 1336   & 1652   & 1693 \\ 
Cut 4        & 133099  & 6392      & 1160   & 1555   & 1654 \\
Cut 5        & 986     & 302       & 285    & 251    & 125 \\
Cut 6        & 422     & 172       & 198    & 86     & 23 \\
Cut 7        & 102     & 65        & 60     & 28     & 8  \\
\hline
Acceptance   & 0.000033 & 0.000014 & 0.0079 & 0.0037 & 0.0011 \\
\hline
\end{tabular}
}
\caption{\small Cut flows of signal and background events in the `1 lepton
+ $\geq$1 b-jet' channel adopted in the CMS experiment. The mass of the top partner is  $m_{tp} =$ 500 GeV. Results of $t_p \bar{t}_p \to$ $bWbW$, $bWth$ and $thth$ are shown separately assuming all $t_p \bar{t}_p$-decays in each mode.}
\label{table: CMS-cutflow}
\end{center}
\end{table}

\begin{table}[t]
\begin{center}
{\small
\begin{tabular}{c|cccccccc}
Mass (GeV) & 400 & 450 & 500 & 550 & 600 & 650 & 700 &750 \\
\hline
$T\bar{T} \to bWbW$ & 0.0025  & 0.0040  & 0.0079 & 0.013  & 0.021  & 0.029
			 & 0.042 & 0.051 \\
$T\bar{T} \to bWth$ & 0.00093  & 0.0020  & 0.0037 & 0.0057 & 0.0094  & 0.014
			 & 0.020 & 0.030 \\
$T\bar{T} \to thth$ & 0.00048 & 0.00065 & 0.0011 & 0.0019 & 0.0031 & 0.0050
                         & 0.0080 & 0.011 \\
\hline
\end{tabular}
}
\caption{\small Acceptance of the signal events in the conventional analysis adopted in the CMS experiment. Masses of the top partner are chosen between 400 GeV and 750 GeV. Results of $t_p \bar{t}_p \to$ $bWbW$, $bWth$ and $thth$ are shown separately assuming all $t_p \bar{t}_p$-decays in each mode.}
\label{table: CMS-acceptance}
\end{center}
\end{table}


\end{document}